\documentclass[twocolumn, journal]{IEEEtran}
\IEEEoverridecommandlockouts

\usepackage{bm}
\usepackage{color}
\usepackage{graphicx}
\usepackage{amsmath}
\usepackage{amssymb}
\usepackage{algorithm}
\usepackage{algorithmic}
\usepackage{ulem}
\usepackage{lineno}
\usepackage{lettrine}
\usepackage{subfigure}
\usepackage{epstopdf}
\usepackage{stfloats}
\usepackage{color}
\usepackage{graphicx}
\usepackage{amsmath}
\usepackage{amssymb}
\usepackage{algorithm}
\usepackage{algorithmic}
\usepackage{amsmath}
\usepackage{multirow}
\usepackage{booktabs}
\usepackage{array}
\usepackage{amsthm}
\usepackage{stfloats}
\usepackage{subfigure}
\usepackage{bm}
\usepackage{booktabs}
\usepackage{setspace}
\usepackage{makecell}
\usepackage{color}

\newcommand{\be}{\begin{equation}}
\newcommand{\ee}{\end{equation}}
\newcommand{\bea}{\begin{eqnarray}}
\newcommand{\eea}{\end{eqnarray}}
\newcommand{\ba}{\begin{array}}
\newcommand{\ea}{\end{array}}
\newcommand{\nid}{\noindent}
\newcommand{\non}{\nonumber}

\title{User Association and Hybrid Beamforming Designs for Cooperative mmWave MIMO Systems
\thanks{Part of this paper has been presented in the IEEE Global  Communications Conference (GLOBECOM), 2021 \cite{glo2021}.}
\thanks{ P. Ni, R. Liu, and M. Li are with the School of Information and Communication
Engineering, Dalian University of Technology, Dalian 116024, China (e-mail:
pfni@mail.dlut.edu.cn; liurang@mail.dlut.edu.cn; mli@dlut.edu.cn).}
\thanks{ Q. Liu is with the School of Computer Science and Technology, Dalian University
of Technology, Dalian 116024, China (e-mail: qianliu@dlut.edu.cn).}
}

\author{Pengfei Ni, Rang Liu,~\IEEEmembership{Graduate Student Member,~IEEE,}\\
        Ming Li,~\IEEEmembership{Senior Member,~IEEE,}
        and Qian Liu,~\IEEEmembership{Member,~IEEE}
}

\begin{document}
\maketitle
\thispagestyle{empty}
\begin{abstract}
% mmWave MIMO hybrid beamforming
Hybrid analog and digital beamforming has emerged as a key enabling technology for millimeter wave (mmWave) massive multiple-input multiple-output (MIMO) communication systems since it can balance the trade-off between system performance and hardware efficiency. Owing to the strong ability of central control, cooperative networks show great potential to enhance the spectral efficiency of mmWave communications.
In this paper, we consider cooperative mmWave MIMO systems and propose user association and hybrid beamforming design algorithms for three typical hybrid beamforming architectures. The central processing unit (CPU) of the cooperative networks first matches the service pairs of base stations (BSs) and users.
Then, an iterative hybrid beamforming design algorithm is proposed to maximize the weighted achievable sum-rate performance of the mmWave MIMO system with fully connected hybrid beamforming architecture. Moreover, a heuristic analog beamforming design algorithm is introduced for the fixed subarray hybrid beamforming architecture. In an effort to further exploit multiple-antenna diversities, we also consider the dynamic subarray architecture and propose a novel antenna design algorithm for the analog beamforming design. Simulation results illustrate that the proposed hybrid beamforming algorithms achieve a significant performance improvement than other existing approaches and the dynamic subarray architecture has great advantages of improving the energy efficiency (EE) performance.
\end{abstract}


\section{Introduction}
%% mmWave + MIMO and supports massive multiple-input multiple-output (MIMO) systems \cite{R. W. Heath}, \cite{E. Bjornson}
To meet the exponentially growing demand of capacity in the next-generation wireless communication systems, millimeter wave (mmWave) communication has been deemed as a promising technology that provides larger available bandwidth. The smaller physical dimension of antenna arrays enables the use of massive multiple-input multiple-output (MIMO) technology to provide highly directional beamforming to overcome the severe pathloss in mmWave channels \cite{R. W. Heath}.
Therefore, the massive MIMO technology (i.e., large-scale antenna arrays) has been considered as the key technology to make mmWave communication from theory to reality.

% practical model
\begin{figure}[t]
\centering
\includegraphics[width= 2.8 in]{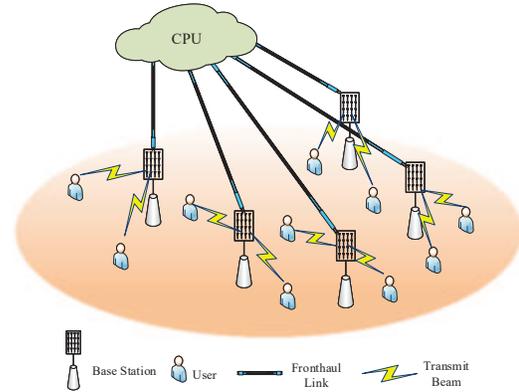}
\vspace{ -0.3 cm}
\caption{Cooperative mmWave MIMO network.}\label{fig:Practical Cooperative network}
\vspace{ -0.3 cm}
\end{figure}

%% Traditional beamforming + hybrid beamforming
Although the fully digital beamforming has been widely employed in traditional communication systems, it requires a radio frequency (RF) chain per antenna element, which leads to significant power consumption, hardware complexity, and cost.
To tackle this problem, hybrid analog and digital beamforming was proposed as an alternative approach for mmWave massive MIMO systems to make a trade-off between performance and cost \cite{X. Zhang2005}.
Recently, three hybrid beamforming architectures have been widely studied in mmWave massive MIMO communication systems, i.e., fully connected architecture, fixed subarray architecture, and dynamic subarray architecture.
% fully connected
In the fully connected hybrid beamforming architecture, each RF chain connects to all antenna elements with a large number of hardware-efficient PSs.
It has been shown in \cite{T. Lin}-\cite{C. Fang} that the fully connected hybrid beamforming architecture can achieve almost the same performance as the fully digital beamforming scheme.
% fixed sub-array
In order to reduce the hardware complexity and power consumption, the fixed subarray hybrid beamforming architecture is suggested, which has fewer PSs by allowing each RF chain connected to a non-overlapping subarray \cite{X. Yu}-\cite{C. Fang2}.
% dynamic sub-array
Furthermore, the researchers also propose dynamic subarray hybrid beamforming architecture to exploit the multiple-antenna diversities and the flexibility of dynamic scheme, in which each RF chain dynamically adjusts the antenna subarray according to channel state information (CSI).
Several literatures have demonstrated the advantages of energy efficiency (EE) performance by applying the dynamic subarray hybrid beamforming architecture to mmWave massive MIMO systems \cite{S. Park}-\cite{L. Yan}.

%% cooperative network
Besides massive MIMO and various hybrid beamforming techniques, changing the structure of mmWave communication networks is also an effective way to improve system performance, e.g., cooperative network and cell-free network \cite{D. Gesbert2010}-\cite{B. He2019}.
As shown in Fig. \ref{fig:Practical Cooperative network}, multiple base stations (BSs) communicate with a central processing unit (CPU) via fronthaul links to realize collaborative communication.
Since the transmitted data are centralized in CPU and shared among all BSs, we can jointly design the power allocation, optimize the beam direction, and coordinate precoding operations to cooperatively manage the inter-cell interference.
Owing to its strong ability of central control and cooperation, cooperative networks have received extensive attention and research in recent years \cite{I. Ahmed}-\cite{Q. Hou}.
One of key components in cooperative network is to determine which user a BS should serve for, i.e., user-BS association \cite{S. Tong}.
In \cite{C. Lee}, the authors optimized the hybrid beamforming and user association in a downlink multi-user cooperative transmission network with two performance metrics: Weighted sum-rate and max-min fairness.
The authors in \cite{J. Jiang} illustrated the significance of user selection in eliminating multi-user interference and balancing the trade-off between hardware complexity and system performance for mmWave massive MIMO systems.
The paper \cite{Q. Hou} studied the optimization of user scheduling and hybrid precoding design in the cloud radio access network (C-RAN).
The authors in \cite{WCNC2021} considered the jointly design of user scheduling and hybrid beamforming to maximize the sum-rate performance.
The user scheduling and analog beamformer were jointly designed and the digital beamformer was optimized by the normalized weighted minimum-mean-square-error (wMMSE) approach.
Motivated by these literatures, we attempt to take both the advantages of hybrid beamforming architectures and cooperative networks.
%The various hybrid beamforming architectures help to reduce the hardware cost and power consumption of mmWave transceivers.
%The cooperative network enables the collaboration of multiple BSs so that the multi-user interference can be reduced, or even eliminated.
%In this paper, we not only consider the design of user association scheme, but also propose a hybrid beamforming algorithm for fully connected architecture to enhance the weighted achievable sum-rate performance in the cooperative mmWave massive MIMO systems.Meanwhile, we also extend the analog beamforming design algorithm to the sub-array architecture, both for the fixed sub-array architecture and dynamic sub-array architecture.

\subsection{Contributions}
In this paper, we investigate the user association and hybrid beamforming designs for three typical antenna architectures in cooperative mmWave MIMO systems.
Our major contributions are summarized as follows:
\begin{itemize}
% 用户调度
\item We consider the user association between users and BSs in the cooperative mmWave MIMO system, which reduces the multi-user interference and improves the system performance. The two-sided stable matching approach is used to realize the successive user association with significantly lower computational complexity than the exhaustive search method.
% 全连接架构
\item With the aid of efficient closed-form fractional programming (FP) theory, we first transform the intractable weighted sum-rate maximization problem into a solvable problem and then propose an iterative hybrid beamforming design algorithm for the fully connected architecture to enhance the sum-rate performance. The analog beamformers of all BSs are optimized only once by our proposed analog beamforming design algorithm and the digital beamformers are obtained with the closed-form solution by the Lagrange multiplier method.
% 部分连接架构
\item In order to further reduce the hardware complexity and power consumption, we provide two algorithms to extend the proposed hybrid beamforming designs into the fixed subarray and dynamic subarray architectures, respectively. K-means based antenna design algorithm is proposed for the dynamic subarray architecture to explore the multiple-antenna diversities and enhance the EE performance.
% 算法性能优势
\item Extensive simulation results are provided to verify and confirm the effectiveness of proposed algorithms in the cooperative mmWave massive MIMO systems. The hybrid beamforming design algorithm for the fully connected structure can achieve a near-optimal performance close to that of the fully digital beamforming. Moreover, with the proposed K-means based antenna design algorithm, the dynamic subarray architecture can achieve higher EE performance than other architectures.
\end{itemize}

\section{System Model and Problem Formulation}

\subsection{System Model}

\begin{figure}[t]
\centering
\includegraphics[width= 3.3 in]{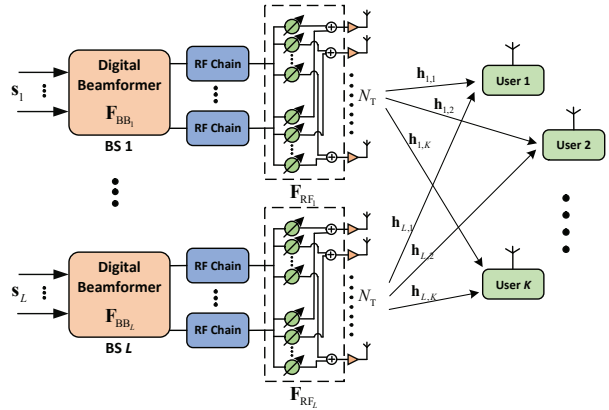}
\vspace{ -0.3 cm}
\caption{Cooperative mmWave MIMO network with fully connected hybrid beamforming architecture.}\label{fig:model}
\vspace{ -0.4 cm}
\end{figure}

We consider the downlink transmissions of a cooperative mmWave massive MIMO system consisting of $L$ BSs and serving $K$ single-antenna users, as shown in Fig. \ref{fig:model}.
Each BS is equipped with $N_{\text{RF}}$ RF chains and $N_{\text{T}}$ antennas, $N_{\text{T}} \gg K$.
The hybrid beamforming scheme is employed and each BS contains a BB processor to perform the digital beamformer and $N_{\text{RF}}$ RF chains to implement the analog beamformer.
Each BS transmits a data stream to one user via one RF chain.
Therefore, the maximum number of users that each BS can support depends on the number of RF chains.
Let $\mathcal{K}_l$ denote the set of users served by the $l$-th BS.
If the number of users served by the $l$-th BS is at most $N_{\text{RF}}$ and each user is served by one BS, i.e., $|\mathcal{K}_l| \leq N_{\text{RF}}$, the total number of users satisfies $K \leq LN_{\text{RF}}$.
In this paper, we adopt the time division duplex (TDD) transmission framework and each coherence interval is divided into three phases: Uplink training, downlink payload data transmission, and uplink payload data transmission.
In the uplink training, users send pilot sequences to BSs. Then, each BS estimates the CSIs of all users and transmits the collected CSIs to CPU.
For the downlink transmission, CPU firstly implements association between $L$ BSs and $K$ users, and then designs the hybrid beamformer with the obtained CSIs.
In order to provide more clearly description, we define the association relationship between BS-$l$ and user-$k$ as matching indicator $\alpha_{l,k} \in \{0, 1\}, l=1, \ldots, L, k = 1, \ldots,K$.
If user-$k$ is associated with BS-$l$, $\alpha_{l,k} = 1$; otherwise, $\alpha_{l,k} = 0$.
Basing on the above analysis, the main signaling overhead involved in considered cooperative network is the exchange of CSIs and the transmission of designed hybrid beamformer between BSs and CPU.
The amount of signaling overhead can be repressed as $LN_{\text{T}}K$ and $L(N_{\text{T}}N_{\text{RF}}+N_{\text{RF}}K)$ for backhaul and fronthaul links, respectively.
With the assumption that the number of antennas $N_{\text{T}}$ is much larger than the number of users $K$, the whole signaling overhead can be expressed
as $LN_{\text{T}}(N_{\text{RF}}+K)$.

\subsection{Subarray Architectures}
% three architectures
\begin{figure}[t]
\centering
\includegraphics[width = 3.5  in]{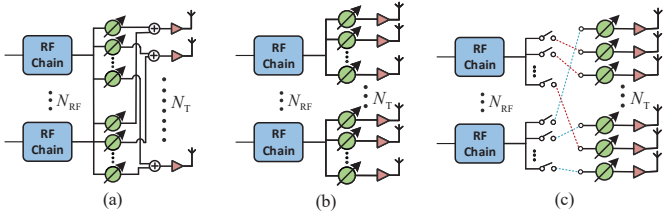}
\vspace{ -0.4 cm}
\caption{Three typical hybrid beamforming architectures: (a) fully connected; (b) fixed subarray; (c) dynamic subarray.}
\vspace{ -0.4 cm}
\label{fig:three architectures}
\end{figure}

As shown in Fig. \ref{fig:model}, the hybrid beamforming architecture in the dotted line square is fully connected antenna array architecture, in which each RF chain is connected to all antennas with a large number of hardware-efficient PSs.
Furthermore, according to the mapping from RF chains to antennas, the hybrid beamforming subarray architectures can be classified into fixed subarray and dynamic subarray architectures, as shown in Fig. \ref{fig:three architectures}(b) and Fig. \ref{fig:three architectures}(c), respectively.
The subarray architecture significantly reduces the power consumption and hardware implementation complexity by connecting RF chains with only a part of antennas.
On the other hand, the dynamic subarray architecture explores multiple-antenna diversities and the flexibility of antenna architecture, which has significant advantages in improving EE performance of mmWave cooperative systems.
Hence, in this paper, we also pay more attention on the hybrid analog and digital beamforming design for the subarray architectures in Sec. V.

\subsection{Downlink Transmission}
As for the hybrid beamforming scheme, the data signal for user-$k$ in the $l$-th BS, denoted by $s_{k}, k =1 , \ldots, K,$ satisfying $\mathbb{E}\{s_ks_k^*\}=1$ and $\mathbb{E}\{s_is_j^*\}=0$, $\forall i\neq j$, is firstly precoded by a digital beamforming vector $\mathbf{f}_{\text{BB}_{l,k}} \in \mathbb{C}^{N_{\text{RF}}}$, and then processed through an analog beamforming matrix $\mathbf{F}_{\text{RF}_l} \triangleq [\mathbf{f}_{\text{RF}_{l,1}}, \ldots, \mathbf{f}_{\text{RF}_{l,N_{\text{RF}}}}] \in \mathbb{C}^{N_{\text{T}} \times N_{\text{RF}}}$, which is implemented in the RF precoder using PS network.
Taking the matching indicator $\alpha_{l,k}$ between BS-$l$ and user-$k$ into account, the precoded signal of BS-$l$ can be represented as
\begin{equation}
\begin{aligned}\label{eq:transmitted signal}
\mathbf{x}_l =  \sum_{k=1}^{K} \alpha_{l,k} \mathbf{F}_{\text{RF}_l} \mathbf{f}_{\text{BB}_{l,k}}s_{k}.
\end{aligned}
\end{equation}

\nid Therefore, the transmit power constraint of the $l$-th BS is
\begin{equation}
\label{eq:power consume}
\begin{aligned}
\sum_{k=1}^{K} \mathrm{Tr}(\alpha_{l,k} \mathbf{f}_{\text{BB}_{l,k}}^H\mathbf{F}_{\text{RF}_l}^H\mathbf{F}_{\text{RF}_l} \mathbf{f}_{\text{BB}_{l,k}}) \leq P_{\text{max}, l},
\end{aligned}
\end{equation}
where $P_{\text{max}, l}$ is the maximum transmit power of BS-$l$.
We assume that all BSs have the same maximum transmit power, i.e., $P_{\textrm{max}}=P_{\textrm{max}, l},\forall l$.

Similar to that in \cite{R. W. Heath}, the mmWave propagation channel is characterized by a multi-path spatial model with $N_{\text{r}}$ rays.
With the assumption that uniform linear array (ULA) is employed with $N_{\text{T}}$ antennas at transmitters and single-antenna at receivers, the channel vector $\mathbf{h}_{l,k} \in \mathbb{C}^{ N_{\text{T}}},\forall l, k,$ is given by
\begin{equation}
\begin{aligned}\label{eq:mmWave channel}
\mathbf{h}_{l,k}=\sqrt{\frac{N_{\text{T}}}{N_{\text{r}}}} \sum_{n=1}^{N_{\text{r}}} \upsilon_{n}\mathbf{a} (\theta_{n}),
\end{aligned}
\end{equation}
where $\upsilon_{n}$ is the complex gain of the $n$-th propagation path and the array response vector $\mathbf{a}(\theta) \in \mathbb{C}^{N_\text{T}}$ can be expressed as
\begin{equation}
\begin{aligned}
\mathbf{a}(\theta)= \frac{1}{\sqrt{N_\text{T}}}[1, {e}^{j\frac{2\pi}{\lambda}d\sin(\theta)}, \ldots , {e}^{j(N_{\text{T}}-1)\frac{2\pi}{\lambda}d\sin(\theta)}]^T ,
\end{aligned}
\end{equation}
where $d$ denotes the distance between antenna elements and $\lambda$ represents the wavelength of mmWave signal.

Thus, the received signal $y_k$ at the $k$-th user is given by
\begin{equation}
\begin{aligned}\label{eq:received signal}
y_k&=\sum_{l=1}^{L} \mathbf{h}^H_{l,k}\mathbf{x}_l+ n_k, \\
&\textcolor{black}{\overset{(a)}{=} \sum_{l=1}^{L} \mathbf{h}^H_{l,k} \sum_{k=1}^{K} \alpha_{l,k}\mathbf{F}_{\text{RF}_l}\mathbf{f}_{\text{BB}_{l,k}}s_{k}+ n_k,}\\
&\textcolor{black}{\overset{(b)}{=} \sum_{l=1}^{L}  \alpha_{l,k}\mathbf{h}^H_{l,k}\mathbf{F}_{\text{RF}_l}\mathbf{f}_{\text{BB}_{l,k}}s_{k}}\\
&~~+\sum_{j =1 ,j\neq k}^{K} \sum_{l=1}^{L} \alpha_{l,j} \mathbf{h}^H_{l,k} \mathbf{F}_{\text{RF}_l} \mathbf{f}_{\text{BB}_{l,j}} s_{j} + n_k,
\end{aligned}
\end{equation}
where $n_k$ denotes the complex additive white Gaussian noise (AWGN) satisfying the complex Gaussian distribution with zero mean and $\sigma_k^2$ variance, i.e., $n_k \sim \mathcal{CN}(0, \sigma^2_k)$, $\forall k$;
$(a)$ is obtained by substituting equation (\ref{eq:transmitted signal}) into the first line of equation (\ref{eq:received signal}), and $(b)$ aims to separately express desired signal and multi-cell multi-user interference for user-$k$.
Similar to existing studies on hybrid beamforming designs (e.g., \cite{Y. Zhang}, \cite{X. Yu}), we assumed that the transmitted signals and CSIs are perfectly known for CPU.
In practice, CSI can be efficiently obtained by existing channel estimation algorithms (e.g., the adaptive compressed sensing (CS) algorithm in \cite{A. Alkhateeb}) at the transmitters and further shared to CPU with the reliable fronthaul links, which can support large signaling overhead transmissions.

Then, the signal to interference-plus-noise ratio (SINR) at the $k$-th user can be easily calculated by
\begin{equation}
\begin{aligned}\label{eq:SINR}
\hspace{-0.1 cm} \mathrm{SINR}_k = \frac {\Big|\sum\limits_{l=1}^{L} \alpha_{l,k}\mathbf{h}_{l,k}^H \mathbf{F}_{\text{RF}_l}\mathbf{f}_{\text{BB}_{l,k}} \Big|^2} {\sum\limits_{j=1,j \neq k}^{K} \Big|\sum\limits_{l=1}^{L} \alpha_{l,j}\mathbf{h}_{l,k}^H \mathbf{F}_{\text{RF}_l}\mathbf{f}_{\text{BB}_{l,j}}\Big|^2 + \sigma_k^2}.
\end{aligned}
\end{equation}

\nid Finally, the weighted achievable sum-rate $R_{\rm{sum}}$ of all $K$ users can be expressed as
\begin{equation}
\begin{aligned}\label{eq:rate}
R_{\rm{sum}} = \sum_{k=1}^{K} \omega_k \mathrm{log}_2(1+\mathrm{SINR}_k),
\end{aligned}
\end{equation}
where weight coefficient $\omega_k \in \mathbb{R}^+$ indicates the priority of user-$k$, $\forall k$.

\subsection{Problem Formulation}
We focus on the weighted sum-rate maximization problem in the cooperative mmWave MIMO systems by optimizing user association matching indicator $\{\alpha_{l,k}\}_{\forall l, \forall k}$ and the hybrid beamformer $\{\mathbf{F}_{\text{RF}_l}, \mathbf{f}_{\text{BB}_{l,k}}\}_{\forall l, \forall k}$ subject to the transmit power constraints at BSs and the unit modulus constraints of PSs.
With the transmit power constraints (\ref{eq:power consume}) and the objective function (\ref{eq:rate}), the weighted sum-rate maximization problem can be formulated as
\begin{subequations}
\label{pr:original}
\begin{align}
\underset{\{\alpha_{l,k},\mathbf{F}_{\text{RF}_l}, \mathbf{f}_{\text{BB}_{l,k}} \}_{\forall l, \forall k}}{\max} &R_{\rm{sum}}\big(\alpha_{l,k},\mathbf{F}_{\text{RF}_l}, \mathbf{f}_{\text{BB}_{l,k}}\big)\\
\mathrm{s.t.}~~~~&\sum_{l=1}^{L}\alpha_{l,k} = 1,~\forall k,\\
&\sum_{k=1}^{K} \alpha_{l,k} = N_{\text{RF}},~\forall l,  \\
&|\mathbf{F}_{\text{RF}_l}(i,j) | = 1,~\forall l,i,j, \\
\sum_{k=1}^{K}\mathrm{Tr}(\alpha_{l,k}& \mathbf{f}_{\text{BB}_{l,k}}^H \mathbf{F}_{\text{RF}_l}^H\mathbf{F}_{\text{RF}_l} \mathbf{f}_{\text{BB}_{l,k}}) \leq P_{\textrm{max}},~\forall l,
\end{align}
\end{subequations}
where $P_{\text{max}}$ is the maximum transmit power for all BSs.
Constraints (8b) and (8c) indicate that each user is served by one BS and each BS can serve at most $N_{\text{RF}}$ users, respectively.
Constraint (8d) represents the constant modulus of PSs and constraint (8e) indicates that the hybrid beamforming matrices cannot be designed to exceed the maximum transmit power.

It is worth noting that there are three reasons and difficulties for optimizing the weighted sum-rate maximization problem (\ref{pr:original}).
First, the form of the objective function $R_{\rm{sum}}$ is not only complicated but also contains fractional terms.
Second, the integer variables $\{\alpha_{l,k}\}_{\forall l, \forall k}$ in (\ref{pr:original}b) and (\ref{pr:original}c) turn the problem (\ref{pr:original}) into a combinatorial problem.
Last, the non-convex unit modulus (\ref{pr:original}d) of analog beamformer $\{\mathbf{F}_{\text{RF}_l}\}_{\forall l}$ is also a tricky constraint to deal with.
Although exhaustive search method can be used to design hybrid beamformer by examining all possible integer variables $\{\alpha_{l,k}\}_{\forall l, \forall k}$, the computational complexity will increase exponentially with the increasing number of users and BSs.
Hence, we adopt a sub-optimal solution \cite{D. Zhao} to efficiently solve problem (\ref{pr:original}) by dividing it into two sub-problems, i.e., user association design and hybrid beamforming design.
Firstly, we use an equivalent metric to effectively design the user association matching indicator $\{\alpha_{l,k}\}_{\forall l, \forall k}$.
Intuitively, the equivalent metric sum-channel-gain not only reduce the computational complexity of optimization, but also achieves sub-optimal performance of original objective function.
Then, the hybrid beamformer $\{\mathbf{F}_{\text{RF}_l}, \mathbf{f}_{\text{BB}_{l,k}}\}_{\forall l,\forall k}$ are optimized by an iterative algorithm.

\section{User Association Design}
\label{sec:PFA for full}
Due to the directionality of mmWave channel, the SINR of each user is heavily dependent on the channel gain.
Thus, we propose to determine the matching indicator $\{\alpha_{l,k}\}_{\forall l, \forall k}$ by maximizing the sum-channel-gain, which is formulated as
\begin{subequations}
\label{pr:user}
\begin{align}
\underset{\{\alpha_{l,k}\}_{\forall l, \forall k}}{\max}~~& \sum_{l=1}^{L}  \sum_{k=1}^{K} \alpha_{l,k}|\mathbf{h}_{l,k}\mathbf{h}_{l,k}^H|^2\\
\mathrm{s.t.}~~~&\sum_{l=1}^{L}  \alpha_{l,k} = 1,~\forall k,  \\
&\sum_{k=1}^{K}  \alpha_{l,k} = N_{\text{RF}},~\forall l.
\end{align}
\end{subequations}
Essentially, problem (\ref{pr:user}) can be posed as a matching problem between BSs and users, which can be solved by the exhaustive search method with $\mathcal{O}(K^L)$ computational complexity \cite{Y. Gu}.
Inspired by the matching theory \cite{Y. Gu}, \cite{Z. Song}, we use the two-sided stable matching approach to realize a successive user association mechanism with lower computational complexity.
Intuitively, the equivalent metric sum-channel-gain ignores interference among users but achieves near-optimal performance with the proposed approach.

As we mentioned before, the SINR or the achievable rate of the $k$-th user is heavily dependent on the channel gain between user-$k$ and BS-$l$, $\forall l$.
Hence, according to different channel states, i.e., channel gains, each BS has a list of preference degrees for service, which can also be interpreted as a descending order of channel gains from BS-$l$ to $K$ users.
In the preference relation ranking (PRR) \cite{Y. Gu}, the ranking from top to bottom indicates the highest to the lowest preferences of users, whom the BSs prefer to serve.
In other words, BS-$l$ intends to serve the first $N_{\textrm{RF}}$ users in the preference relation ranking.
Since unilateral selection cannot represent the optimal double-link between users and BSs, the relationship between users and BSs must be two-sided selection.
Similarly, each user also has a list of preference relations of BSs to access.
Thus, if the $l$-th BS is the first one in the $k$-th user's preference relation ranking and the $k$-th user is also the most preferred user that the $l$-th BS wants to serve, then set $\alpha_{l,k} = 1$.
Specifically, the two-sided stable matching approach of user association is described in the following four steps.

\newcounter{TempEqCnt} % 创建临时变量TempEqCnt
\setcounter{TempEqCnt}{\value{equation}} % 将当前公式序号 赋给TempEqCnt
\setcounter{equation}{13} % 当前公式序号变为x，x等于长公式应有的序号减1.
\begin{figure*}[!t] %hb代表放在文章底部，%ht为放在文章顶部
\begin{equation}
\begin{aligned}
\label{eq:fksee}
f_{o}(\bm{\rho}, \bm{\xi},\mathbf{F}_{\text{RF}_l}, \mathbf{f}_{\text{BB}_{l,k}})=&\sum_{k=1}^{K}\omega_k \text{log}_2(1+\rho_k)-\sum_{k=1}^{K} \omega_k \rho_k +\sum_{k=1}^{K} 2 \sqrt{\omega_k(1+\rho_k)} \Re\{ \xi_k^* \sum_{l=1}^{L} \alpha_{l,k}\mathbf{h}_{l,k}^H \mathbf{F}_{\text{RF}_l}\mathbf{f}_{\text{BB}_{l,k}} \}\\
&- \sum_{k=1}^{K} \omega_k |\xi_k|^2 \sum_{l=1}^{L} \sum_{j=1}^{K}  \alpha_{l,j}|\mathbf{h}_{l,k}^H \mathbf{F}_{\text{RF}_l}\mathbf{f}_{\text{BB}_{l,j}}|^2 + \sum_{k=1}^{K} \omega_k|\xi_k|^2\sigma_k^2.
\end{aligned}
\end{equation}
\centering{\rule[-2 pt]{18.2cm}{0.05em}}
\end{figure*}

\vspace{0.5 cm}
\hangafter 1
\setlength{\hangindent}{3.3em}
\nid \textit{Step 1:} Initialize $\mathcal{K} := \{1, \ldots, K\}$, $\mathcal{L} := \{1, \ldots, L\}$, and $\alpha_{l,k} = 0$, $\forall l, \forall k$, respectively. Calculate the channel gains of all BSs and users, and obtain $(L+K)$ PRRs. Next, select the best BS-$l^\star$ for user-$k$, which is the first BS of user-$k$'s PRR:
\setcounter{TempEqCnt}{\value{equation}} % 将当前公式序号 赋给TempEqCnt
\setcounter{equation}{9} % 当前公式序号变为x，x等于长公式应有的序号减1.
\bea
l^\star = \mathrm{arg}~\underset{l \in \mathcal{L}}{\max} ~|\mathbf{h}_{l,k}\mathbf{h}_{l,k}^H|^2.
\eea

\hangafter 1
\setlength{\hangindent}{3.3em}
\nid \textit{Step 2:} If the condition $\sum_{k \in \mathcal{K}}\alpha_{l^{\star},k} = N_{\text{RF}}$ has been satisfied, let $\mathcal{L} := \mathcal{L} - l^\star$, remove BS-$l^\star$ from the $k$-th user's PRR, and go back to Step 1. Otherwise, select the best user-$k_0^\star$, which is the first user of the $l^\star$-th BS's PRR:
\bea
k_0^\star =\mathrm{arg}~\underset{k \in \mathcal{K}}{\max}~ |\mathbf{h}_{l^\star,k} \mathbf{h}_{l^\star,k}^H|^2.
\eea

\hangafter 1
\setlength{\hangindent}{3.3em}
\nid \textit{Step 3:} If $k_0^\star = k$, then set $\alpha_{l^\star,k} = 1$ and update the user set as $\mathcal{K}: = \mathcal{K} - k_0^\star$.
Otherwise, remove BS-$l^\star$ from the $k$-th user's PRR, and go back to \textit{Step 1}.

\hangafter 1
\setlength{\hangindent}{3.3em}
\nid \textit{Step 4:} Repeat \textit{Steps 1-3} until $\mathcal{K} = \varnothing$, $\mathcal{L} = \varnothing$.
\vspace{0.5 cm}

After \textit{Steps 1-4}, the BS-user matching indicators $\{\alpha_{l,k}\}_{\forall l, \forall k}$, can be obtained with significantly lower computational complexity than the exhaustive search method, which has $\mathcal{O}(K^L)$ computational complexity.
The two-sided stable matching approach has at most $K$ (i.e., $\min\{K, LN_{\textrm{RF}}\}$) iterations \cite{Y. Zhang2021}.
Moreover, as for finding the best BS with maximum channel gain for user-$k$ and selecting the best user for BS-$l$, the sorting algorithm is required, which has $\mathcal{O}(n\log n)$ computational complexity \cite{T. H. Cormen sort}.
Then, the computational complexities of \textit{Step 1} and \textit{Step 2} are $\mathcal{O}(L \log L)$ and $\mathcal{O}(K \log K )$, respectively.
Hence, the total computational complexity for the two-sided stable matching approach is $\mathcal{O}\big(K(L \log L+K \log K)\big)$.

\vspace{-0.2cm}
\section{Hybrid Beamforming Design For Fully Connected Architecture}
\label{sec:PFA}
In this section, the hybrid beamforming design algorithm for fully connected architecture is introduced.
Firstly, we assume that the user association design has been completed and the matching indicators have been known for CPU before the hybrid beamforming design.
Next, we use the much efficient closed-form FP approach to deal with the complex form of objective function and then propose an iterative hybrid beamforming design algorithm to optimize the introduced variables, analog beamforming matrix, and digital beamformer until convergence condition is met.

\vspace{-0.4cm}
\subsection{Closed-form FP-aided Transformation}
\label{sec:PFA for digital}
With the user association matching indicator $\{\alpha_{l,k}\}_{\forall l, \forall k}$ obtained in Sec. III, the fully connected hybrid beamforming design problem is formulated as follows
\begin{subequations}\label{pr:hybrid beamforming for fc}
\begin{align}
\underset{\{\mathbf{F}_{\text{RF}_l}, \mathbf{f}_{\text{BB}_{l,k}}\}_{\forall l, \forall k}  }{\max} &R_{\rm{sum}}(\mathbf{F}_{\text{RF}_l}, \mathbf{f}_{\text{BB}_{l,k}} )\\
\mathrm{s.t.}~~~~~&|\mathbf{F}_{\text{RF}_l}(i,j) | = 1,~\forall l,i,j, \\
\sum_{k=1}^{K}\mathrm{Tr}(\alpha_{l,k}& \mathbf{f}_{\text{BB}_{l,k}}^H \mathbf{F}_{\text{RF}_l}^H\mathbf{F}_{\text{RF}_l} \mathbf{f}_{\text{BB}_{l,k}}) \leq P_{\textrm{max}},~\forall l.
\end{align}
\end{subequations}

\nid Obviously, the main challenge of designing hybrid beamforming matrices in problem (\ref{pr:hybrid beamforming for fc}) is dealing with complex objective function (\ref{pr:hybrid beamforming for fc}a) and non-convex constraints (\ref{pr:hybrid beamforming for fc}b).
Motivated by \cite{K. Shen1}, \cite{K. Shen2}, both the direct FP and the closed-form FP approaches can be used to solve the considered optimization problem.
The direct FP method only applies the \textit{Quadratic Transform} to convert the SINR terms (i.e., fractional terms) inside the log function into a concave function of $\{\mathbf{F}_{\text{RF}_l}\mathbf{f}_{\text{BB}_{l,k}}\}_{\forall l, \forall k}$.
When the introduced variables are held fixed, a convex problem with variables $\{\mathbf{F}_{\text{RF}_l} \mathbf{f}_{\text{BB}_{l,k}}\}_{\forall l,\forall k}$ (or two convex problems with analog beamformers $\{\mathbf{F}_{\text{RF}_l}\}_{\forall l}$ and digital beamformers $\{\mathbf{f}_{\text{BB}_{l,k}}\}_{\forall l,\forall k}$, respectively) is needed to optimize by convex optimization tools (e.g., CVX) with higher computational complexity.
However, how to split the analog beamforming matrices and digital beamformers from $\{\mathbf{F}_{\text{RF}_l} \mathbf{f}_{\text{BB}_{l,k}}\}_{\forall l,\forall k}$ is still a thorny issue.
As for the closed-form FP approach, by introducing two more variables, the sum-logarithm problem is first reformulated into a sum-of-ratio form with \textit{Lagrangian Dual Transform}, and then \textit{Quadratic Transform} is applied to fractional terms to represent all variables with a differentiable objective function.
Because the closed-form FP approach updates all variables in the closed-form while the direct FP requires a convex optimization in each iteration, the closed-form FP approach is more efficient on a per-iteration basis.
Therefore, we employ the much preferred closed-form FP approach to transform the unsolvable objective function into an equivalent form, which can be differentiated with respect to all variables.

By introducing two auxiliary vectors $\bm{\rho} \triangleq[\rho_1, \ldots, \rho_k]$ and $\bm{\xi} \triangleq[\xi_1, \ldots, \xi_k]$, the original problem (\ref{pr:hybrid beamforming for fc}) is reformulated as
\vspace{-0.15 cm}
\begin{subequations}\label{pr:fourvariables}
\begin{align}
\underset{\{\bm{\rho}, \bm{\xi},\mathbf{F}_{\text{RF}_l}, \mathbf{f}_{\text{BB}_{l,k}}\}_{\forall l, \forall k} } {\max}& f_{o}(\bm{\rho}, \bm{\xi},\mathbf{F}_{\text{RF}_l}, \mathbf{f}_{\text{BB}_{l,k}} ) \\
\mathrm{s.t.}~~~~~~&(\ref{pr:hybrid beamforming for fc}\text{b}), (\ref{pr:hybrid beamforming for fc}\text{c}),
\end{align}
\end{subequations}
where $f_{o}(\bm{\rho}, \bm{\xi},\mathbf{F}_{\text{RF}_l}, \mathbf{f}_{\text{BB}_{l,k}} )$ is the transformed objective function with the aid of closed-form FP approach and it has been shown at the top of this page.
As we mentioned before, all variables can be updated in the closed-form.
Thus, we propose an iterative hybrid beamforming design algorithm (similar to the block coordinate descent (BCD) method) to effectively solve problem (\ref{pr:fourvariables}).
The details of updating each variable are described in the following subsections.

\vspace{-0.4cm}
\subsection{Auxiliary Variable Design}
\label{sec:PFA for digital}
When other variables are held fixed (i.e., $\bm{\xi}, \mathbf{F}_{\text{RF}_l},\mathbf{f}_{\text{BB}_{l,k}} $), the objective function $f_{o}(\bm{\rho}, \bm{\xi},\mathbf{F}_{\text{RF}_l}, \mathbf{f}_{\text{BB}_{l,k}} )$ is a concave function with respect to $\mathbf{\rho}_k,\forall k$.
Therefore, the closed-form solution of $\rho_k$ can be calculated by setting $\partial f_{o} / \partial \mathbf{\rho}_k$ to zero
\setcounter{TempEqCnt}{\value{equation}} % 将当前公式序号 赋给TempEqCnt
\setcounter{equation}{14} % 当前公式序号变为x，x等于长公式应有的序号减1.
\begin{equation}
\begin{aligned}\label{optimal rho}
\mathbf{\rho}_k^{\star} &= \frac{|\sum_{l=1}^{L} \alpha_{l,k}\mathbf{h}_{l,k}^H \mathbf{F}_{\text{RF}_l}\mathbf{f}_{\text{BB}_{l,k}}|^2}{ \sum_{l=1}^{L} \sum_{j=1,j \neq k}^{K} \alpha_{l,j}|\mathbf{h}_{l,k}^H \mathbf{F}_{\text{RF}_l}\mathbf{f}_{\text{BB}_{l,j}}|^2 + \sigma_k^2}.
\end{aligned}
\end{equation}

With fixed $\bm{\rho}, \mathbf{F}_{\text{RF}_l},$ and $\mathbf{f}_{\text{BB}_{l,k}}$, by setting the derivative of $f_{o}$ with respect to $\mathbf{\xi}_k$ to zero, the closed-form solution of $\xi_k$ can be given by
\begin{equation}
\begin{aligned}\label{optimal xi}
\mathbf{\xi}_k^{\star} &= \frac{\sqrt{\omega_k(1+\rho_k)}\sum_{l=1}^{L} \alpha_{l,k}\mathbf{h}_{l,k}^H \mathbf{F}_{\text{RF}_l}\mathbf{f}_{\text{BB}_{l,k}}}{\sum_{l=1}^{L} \sum_{j=1}^{K} \alpha_{l,j}|\mathbf{h}_{l,k}^H \mathbf{F}_{\text{RF}_l}\mathbf{f}_{\text{BB}_{l,j}}|^2+ \sigma_k^2}.
\end{aligned}
\end{equation}

\subsection{Analog Beamformer Design}
\label{sec:analog for fc}
In this subsection, we focus on the optimization of analog beamformer $\mathbf{F}_{\text{RF}_l},\forall l$.
When the auxiliary vectors $\bm{\rho}$, $\bm{\xi}$, and the digital beamformer $\mathbf{f}_{\text{BB}_{l,k}},\forall l,k,$ are fixed, we first recast the objective function $f_{o}(\bm{\rho}^{\star}, \bm{\xi}^{\star},\mathbf{F}_{\text{RF}_l}, \mathbf{f}_{\text{BB}_{l,k}}^{\star})$ by sorting out constant terms and then removing them.
The reformulated objective function $f_{a}(\mathbf{F}_{\text{RF}_l})$ is defined as
\begin{equation}
\begin{aligned}
\hspace{-1.0 em}f_{a}(\mathbf{F}_{\text{RF}_l}) =& ~\tilde{f}_{a}(\mathbf{F}_{\text{RF}_l}) +c,\\ \non
=&\sum_{k = 1}^{K} \hspace{-0.2em} 2 \sqrt{\omega_k(1+\rho_k)}\Re\{ \xi_k^* \sum_{l = 1}^{L}  \alpha_{l,k} \mathbf{h}_{l,k}^H \mathbf{F}_{\text{RF}_l}\mathbf{f}_{\text{BB}_{l,k}} \} \non
\end{aligned}
\end{equation}
\begin{equation}\label{eq:refun of analog}
\begin{aligned}
\hspace{3.3 em} - \sum_{k = 1}^{K} \omega_k |\xi_k|^2 \hspace{-0.2em}\sum_{l = 1}^{L} \hspace{-0.2em} \sum_{j = 1}^{K} \hspace{-0.2em} \alpha_{l,j}|\mathbf{h}_{l,k}^H \mathbf{F}_{\text{RF}_l}\mathbf{f}_{\text{BB}_{l,j}}|^2 + c,
\end{aligned}
\end{equation}
where $c \triangleq \sum_{k = 1}^{K} \omega_k \left(\text{log}_2(1+\rho_k)-\rho_k -|\xi_k|^2 \sigma_k^2\right)$ is the integrated constant term.
Removing the constant term $c$ of reformulated objective function (\ref{eq:refun of analog}), the sub-problem of analog beamforming design can be expressed as
\begin{subequations}\label{pr:analog 1}
\begin{align}
\underset{\{\mathbf{F}_{\text{RF}_l}\}_{\forall l}} {\max}&~~~\tilde{f}_{a}(\mathbf{F}_{\text{RF}_l}) \\
\mathrm{s.t.}~&~~~| \mathbf{F}_{\text{RF}_l}(i,j) | = 1,~\forall l,i,j.
\end{align}
\end{subequations}
\nid Although the transmit power constraints (\ref{pr:hybrid beamforming for fc}\text{c}) are omitted in sub-problem (\ref{pr:analog 1}), it can be satisfied by digital beamforming design in the next subsection.
There are two main difficulties to solve for sub-problem (\ref{pr:analog 1}).
On the one hand, compared to the system with only one BS, $L$ variables (i.e., $\mathbf{F}_{\text{RF}_l}, \forall l$) are optimized in sub-problem (\ref{pr:analog 1}).
On the other hand, the unit modulus constraint (\ref{pr:analog 1}b) is non-convex, which makes problem (\ref{pr:analog 1}) intractable.

To tackle the first issue, we attempt to vectorize the $L$ analog beamforming matrices $\mathbf{F}_{\text{RF}_l}, \forall l$, and then combine them into an intermediate variable  so that we can directly optimize the analog beamformer for all BSs instead of solving it for each BS individually.
Next, we provide the detailed transformation and derivation.
The analog beamformer $\mathbf{F}_{\text{RF}_l}$ are vectorized as $\mathrm{vec}(\mathbf{F}_{\text{RF}_l})$ for all $L$ BSs and the objective function $\tilde{f}_{a}(\mathbf{F}_{\text{RF}_l})$ can be rewritten as
\begin{equation}
\begin{aligned}\label{eq:varphi vec}
\hspace{-0.8em}\tilde{f}_{a1}&(\mathrm{vec}(\mathbf{F}_{\text{RF}_l})^*)=\sum_{l = 1}^{L} 2 \Re \{ (\mathrm{vec}(\mathbf{F}_{\text{RF}_l})^*)^H\tilde{\mathbf{t}}_{l}\}\\
&~~~~~~~~~~~~~~~~~~~~- \sum_{l = 1}^{L}  (\mathrm{vec}(\mathbf{F}_{\text{RF}_l})^*)^H \tilde{\mathbf{T}}_{l}\mathrm{vec}(\mathbf{F}_{\text{RF}_l})^*,
\end{aligned}
\end{equation}
where
\begin{subequations}
\begin{align}
\tilde{\mathbf{t}}_{l}& \triangleq \sum_{k = 1}^{K}\sqrt{\omega_k(1+\rho_k)}\xi_k^* \alpha_{l,k}(\mathbf{I}_{N_{\textrm{RF}}} \otimes \mathbf{h}_{l,k}^H)^T\mathbf{f}_{\text{BB}_{l,k}},\\
\tilde{\mathbf{T}}_{l} & \triangleq \sum_{k = 1}^{K} \hspace{-0.3em} \omega_k |\xi_k|^2 \hspace{-0.2em} \sum_{j = 1}^{K} \hspace{-0.3em} \alpha_{l,j} (\mathbf{I}_{N_{\textrm{RF}}} \hspace{-0.3em} \otimes \hspace{-0.2em}  \mathbf{h}_{l,k}^H)^T\mathbf{f}_{\text{BB}_{l,j}}\mathbf{f}_{\text{BB}_{l,j}}^H(\mathbf{I}_{N_{\textrm{RF}}}\hspace{-0.3em}  \otimes \hspace{-0.2em} \mathbf{h}_{l,k}^H)^*.
\end{align}
\end{subequations}

\nid Secondly, benefit from vectorization, we use an intermediate variable $\mathbf{f} \in \mathbb{C}^{LN_{\textrm{T}}N_{\textrm{RF}}}$ to combine $L$ analog beamformer vectors $\mathrm{vec}(\mathbf{F}_{\text{RF}_l})^* \in \mathbb{C}^{N_{\textrm{T}}N_{\textrm{RF}}}, \forall l$.
By introducing a transformation matrix $\mathbf{\Xi}_l, \forall l$, which has all zeros elements except for the identity matrix $\mathbf{I}_{N_{\textrm{T}}N_{\textrm{RF}}}$ in the rows from $(l-1)N_{\textrm{T}}N_{\textrm{RF}}+1$ to $lN_{\textrm{T}}N_{\textrm{RF}}$, the transformation relationship between vectors $\mathrm{vec}(\mathbf{F}_{\text{RF}_l})^* $ and $\mathbf{f}$ can be expressed as
\begin{equation}\label{fun:f}
\begin{aligned}
\mathrm{vec}(\mathbf{F}_{\text{RF}_l})^* &= \mathbf{\Xi}_l^H\mathbf{f},\\
\mathbf{f} \triangleq[(\mathrm{vec}(\mathbf{F}_{\text{RF}_1})^*)^T&, \ldots, (\mathrm{vec}(\mathbf{F}_{\text{RF}_l})^*)^T]^T.
\end{aligned}
\end{equation}
Then, we substitute equation (\ref{fun:f}) into function (\ref{eq:varphi vec}) and obtain a new function with respect to vector $\mathbf{f}$ as
\begin{equation}\label{fun:varphi f}
\begin{aligned}
f_{a2}(\mathbf{f})= 2\Re \{ \mathbf{f}^H\mathbf{v}\}- \mathbf{f}^H\mathbf{W}\mathbf{f},
\end{aligned}
\end{equation}
where we define two matrices $\mathbf{v} \triangleq \sum_{l = 1}^{L} \mathbf{\Xi}_l\tilde{\mathbf{t}}_{l}$ and $\mathbf{W} \triangleq \sum_{l = 1}^{L}  \mathbf{\Xi}_l \tilde{\mathbf{T}}_{l }\mathbf{\Xi}_l^H$ for brevity.
Hence, the original sub-problem (\ref{pr:analog 1}) can be recast as
\begin{subequations}\label{pr:analog 4}
\begin{align}
\underset{ \mathbf{f}} {\min} &~~\tilde{f}_{a2}(\mathbf{f}) \\
\mathrm{s.t.}&~~| \mathbf{f}(i)| = 1,~\forall i,
\end{align}
\end{subequations}
where $\tilde{f}_{a2}(\mathbf{f}) \triangleq -f_{a2}(\mathbf{f})$.
Up to now, the first issue (i.e., $L$ variables are optimized) has been tackled and only vector $\mathbf{f}$ is optimized in the new reformulated problem (\ref{pr:analog 4}).

Next, as for the designing of vector $\mathbf{f}$, the non-convex unit modulus constraint (\ref{pr:analog 4}b) makes the recast problem (\ref{pr:analog 4}) challenging, which is also the second difficulty that we mentioned before.
We have reformulated the subproblem of analog beamformer design as a general complex-valued homogeneous quadratically constrained quadratic program (QCQP) problem, which can be solved by semidefinite relaxation (SDR) technique \cite{Z. Luo}.
Due to the relaxation of rank-1 constraint, the obtained result of SDR is not an optimal solution in general.
Meanwhile, the interior-point algorithm of SDR has a worst case complexity of $\mathcal{O}\big((LN_{\textrm{T}}N_{\textrm{RF}})^{4.5}\log(1/\epsilon)\big)$, wherein $\epsilon > 0$ is the solution accuracy \cite{Z. Luo}.
Besides, the problem (\ref{pr:analog 4}) can also be tackled by elements-wise BCD method.
As the number of antennas increases, the number of iterations increases significantly, which in turn leads to a slow convergence rate.
Fortunately, the manifold algorithm is an effective approach to solve problem with this typical non-convex constraints \cite{X. Yu}.
Therefore, in the following, we apply the conjugate gradient algorithm on the Riemannian manifold space (also referred to Riemannian conjugate gradient (RCG) algorithm) and propose the RCG-based analog beamformer design algorithm to solve sub-problem (\ref{pr:analog 4}).

Firstly, we consider constraint (\ref{pr:analog 4}b) as a $(LN_{\text{T}}N_{\text{RF}})$-dimensional smooth Riemannian manifold
\begin{equation}
\begin{aligned}
\mathcal{M} = \{ \mathbf{f} \in \mathbb{C}^{LN_{\textrm{T}}N_{\textrm{RF}}}: |\mathbf{f}(i)|=1, \forall i\},
\end{aligned}
\end{equation}
and the corresponding tangent space is
\begin{equation}
\begin{aligned}
\textit{T}_{\mathbf{f}}\mathcal{M} = \{  \mathbf{g} \in \mathbb{C}^{LN_{\text{T}}N_{\text{RF}}}: \mathfrak{R}\{\mathbf{g}(i) \circ \mathbf{f}(i)^\ast \} = 0, \forall i\}.
\end{aligned}
\end{equation}
\nid To implement the conjugate gradient algorithm, we need the Euclidean gradient to determine the corresponding Riemannian gradient and the Euclidean gradient is
\begin{equation}\label{Euclidean gradient}
\begin{aligned}
\nabla \tilde{f}_{a2}(\mathbf{f}) &=(\mathbf{W}+\mathbf{W}^H)\mathbf{f}- 2\mathbf{v}.
\end{aligned}
\end{equation}
Moreover, we define the Riemannian gradient as the tangent vector $\rm{grad}_{\mathbf{f}} \tilde{f}_{a2}(\mathbf{f})$, which can be given by
\begin{equation}
\begin{aligned}
\label{Riemannian gradient}
\mathrm{grad}_{\mathbf{f}} \tilde{f}_{a2}(\mathbf{f}) &= \mathrm{Proj}_{\mathbf{f}}\{ \nabla \tilde{f}_{a2}(\mathbf{f})\},\\~
&= \nabla \tilde{f}_{a2}(\mathbf{f}) - \mathfrak{R}\{ \nabla \tilde{f}_{a2}(\mathbf{f}) \circ \mathbf{f}^*\}\circ \mathbf{f},
\end{aligned}
\end{equation}
where $\mathrm{Proj}_{\mathbf{f}}\{\cdot\}$ denotes the orthogonal projection of $\nabla  \tilde{f}_{a2}(\mathbf{f})$ onto the tangent space $T_{\mathbf{f}}\mathcal{M}$.
% 共轭梯度算法计算
Then, we apply the fast linear search conjugate gradient algorithm into Riemannian space to optimize vector $\mathbf{f}$.
Due to the characteristics of Riemannian space, the traditional linear search method of conjugate gradient algorithm cannot be directly applied to Riemannian space.
In the $i$-th iteration, the search direction $\mathbf{d}^{(i)}$ is calculated by the Riemannian gradient $\mathrm{grad}_{\mathbf{f}} \tilde{f}_{a2}(\mathbf{f}^{(i)})$ and the ($i$-1)-th search direction $\mathbf{d}^{(i-1)}$.
Since these two vectors are located in different tangent spaces, they cannot be added together directly.
Thus, the concept of $\textit{vector transport}$ is introduced to transport $\mathbf{d}^{(i-1)}$ into the tangent space of Riemannian gradient $\mathrm{grad}_{\mathbf{f}} \tilde{f}_{a2}(\mathbf{f}^{(i)})$.
Thus, the search direction in the $i$-th iteration can be determined as
\begin{equation}
\begin{aligned}\label{Search direction}
\mathbf{d}^{(i)}= -\mathrm{grad}_{\mathbf{f}} \tilde{f}_{a2}(\mathbf{f}^{(i)}) + a^{(i)} \mathbf{d}^{(i-1)}_t,
\end{aligned}
\end{equation}
where $a^{(i)}$ is Polak-Ribiere parameter \cite{L. Zhang} and the subscript ``$t$" denotes the $\textit{vector transport}$ operation.
With the obtained search direction $\mathbf{d}^{(i)}$ and $retraction$ operation, the vector $\mathbf{f}$ of the $i$-th updating can be given by
\begin{equation}
\begin{aligned}\label{Retraction}
\mathbf{f}^{(i)}= \mathrm{Ret}_{\mathbf{f}}\{\mathbf{f}^{(i-1)} + b^{(i)}\mathbf{d}^{(i)}\},
\end{aligned}
\end{equation}
where $\mathrm{Ret}_{\mathbf{f}}\{\cdot\}$ represents the $retraction$ operation and $b^{(i)}$ indicates the Armijo backtracking line search step size \cite{L. Zhang}.
With (\ref{Search direction}) and (\ref{Retraction}), the optimal solution of $\mathbf{f}^{\star}$ can be optimized by the RCG-based analog beamformer design algorithm.
Note that the convergency of RCG algorithm has been proved by Theorem 4.3.1 in \cite{P.-A. Absil} and the computational complexity of obtaining the intermediate
variable $\mathbf{f} \in \mathbb{C}^{L N_{\textrm{T}}N_{\textrm{RF}}}$ is approximated by $\mathcal{O}\big((L N_{\textrm{T}}N_{\textrm{RF}})^{1.5}\big)$ \cite{P.-A. Absil}.

% 重构
Lastly, the analog beamformer $\mathbf{F}_{\textrm{RF}_l}^{\star},\forall l,$ for all $L$ BSs can be selected and reconstructed from $\mathbf{f}^{\star}$ through
\begin{equation}
\begin{aligned}\label{select analog for fully}
\hspace{-0.2 cm}\mathbf{F}_{\textrm{RF}_l}^{\star} =\mathrm{reshape} \left(\big(\mathbf{f^{\star}}((l\text{-}1)N\text{+}1:lN)\big)^*, N_{\textrm{T}}, N_{\textrm{RF}}\right),
\end{aligned}
\end{equation}
where $N \triangleq N_{\textrm{T}}N_{\textrm{RF}}$.
In summary, the RCG-based analog beamformer design algorithm to obtain $\mathbf{F}_{\textrm{RF}_l}^{\star}, \forall l$, is summarized in Algorithm 1.

\begin{algorithm}[!t]\small% 算法
  \caption{RCG-based Analog Beamformer Algorithm}
  \label{alg:Algorithm 1}
  \begin{algorithmic}[1]
    \REQUIRE $L, N_{\textrm{T}}, N_{\textrm{RF}}, I_{\mathrm{max}}, \epsilon_{\mathrm{th}}, \tilde{f}_{a2}(\mathbf{f}), \mathbf{f}^0 \in \mathcal{M}$.
    \ENSURE  $\mathbf{F}_{\text{RF}_l}^{\star},  l=1, \ldots, L$.
    \STATE {Initialize $i= 0$, $\mathbf{d}^0 = -\mathrm{grad}_{\mathbf{f}} \tilde{f}_{a2}(\mathbf{f}^0)$.}
    \WHILE {$i < I_{\mathrm{max}}$ and $\epsilon \geq \epsilon_{\mathrm{th}}$}
    \STATE {Calculate Riemannian gradient $\mathrm{grad}_{\mathbf{f}} \tilde{f}_{a2}(\mathbf{f}^{(i)})$ with (\ref{Riemannian gradient});}
    \STATE {Select Polak-Ribiere parameter $a^{(i)}$;}
    \STATE {Calculate search direction $\mathbf{d}^{(i)}$ by (\ref{Search direction});}
    \STATE {Choose Armijo backtracking line search step size $b^{(i)}$;}
    \STATE {Calculate $\mathbf{f}^{(i+1)}$ by (\ref{Retraction});}
    \STATE {$\epsilon  = \| \mathrm{grad}_{\mathbf{f}} \tilde{f}_{a2}(\mathbf{f}^{(i)})\|^2$;}
    \STATE {$i = i + 1$.}
    \ENDWHILE
    \STATE Construct $\mathbf{F}_{\textrm{RF}_l}^{\star},  l=1, \ldots, L,$ by (\ref{select analog for fully}).
  \end{algorithmic}
\end{algorithm}

\subsection{Digital Beamformer Design}
In this subsection, we focus on the optimization of digital beamformer $\{\mathbf{f}_{\textrm{BB}_{l,k}}\}_{\forall l, \forall k}$.
When the other variables are optimized and held fixed, the sub-problem of digital beamforming design can be given by
\begin{subequations}\label{pr:digital 1}
\begin{align}
\underset{\{\mathbf{f}_{\text{BB}_{l,k}}\}_{\forall l, \forall k}} {\max}&f_{o}(\bm{\rho}^{\star}, \bm{\xi}^{\star},\mathbf{F}_{\text{RF}_l}^{\star}, \mathbf{f}_{\text{BB}_{l,k}}) \\
\mathrm{s.t.}~~~~&\sum_{k = 1}^{K} \mathrm{Tr}(\alpha_{l,k} \mathbf{f}_{\text{BB}_{l,k}}^H \mathbf{F}_l^{\star}  \mathbf{f}_{\text{BB}_{l,k}}) \leq P_{\textrm{max}},~\forall l,
\end{align}
\end{subequations}
where we define $\mathbf{F}_l^{\star} \triangleq (\mathbf{F}_{\text{RF}_l}^{\star})^H \mathbf{F}_{\text{RF}_l}^{\star}, \forall l,$ for brevity.
Similar to Sec. IV-C, we also first rewrite the objective function $f_{o}(\bm{\rho}^{\star}, \bm{\xi}^{\star},\mathbf{F}_{\text{RF}_l}^{\star}, \mathbf{f}_{\text{BB}_{l,k}})$ by removing the constant term.
By defining matrices $\mathbf{\Gamma}_l\triangleq \sum_{m=1}^{K} |\xi_m|^2  (\mathbf{F}_{\textrm{RF}_l}^{\star})^H\mathbf{h}_{l,m} \mathbf{h}_{l,m}^H \mathbf{F}_{\textrm{RF}_l}^{\star}$ for all BSs, the new objective function is given by
\begin{equation}
\begin{aligned}
\hspace{ -2.8 cm}f_{d}(\mathbf{f}_{\text{BB}_{l,k}})= \sum_{l =1}^{L} \sum_{k = 1}^{K} \omega_k\alpha_{l,k}\mathbf{f}_{\textrm{BB}_{l,k}}^H \mathbf{\Gamma}_l \mathbf{f}_{\textrm{BB}_{l,k}}
\end{aligned}
\end{equation}
\vspace{ -0.08 cm}
\begin{equation}
\begin{aligned}
\hspace{ 1.5 cm}-\sum_{l =1}^{L} \sum_{k = 1}^{K} 2\sqrt{\omega_k(1+\rho_k)} \Re\{ \xi_k^* \alpha_{l,k} \mathbf{h}_{l,k}^H \mathbf{F}_{\textrm{RF}_l}^{\star} \mathbf{f}_{\textrm{BB}_{l,k}} \}. \non
\end{aligned}
\end{equation}
Then, the sub-problem can be reformulated as
\begin{subequations}\label{pr:digital 2}
\begin{align}
\underset{\{\mathbf{f}_{\text{BB}_{l,k}}\}_{\forall l, \forall k}} {\max} &f_{d}(\mathbf{f}_{\text{BB}_{l,k}})\\
\mathrm{s.t.}~~~~&\sum_{k = 1}^{K} \mathrm{Tr}(\alpha_{l,k} \mathbf{f}_{\text{BB}_{l,k}}^H\mathbf{F}_l^{\star} \mathbf{f}_{\text{BB}_{l,k}}) \leq P_{\textrm{max}},~\forall l.
\end{align}
\end{subequations}
Since the matrices $\mathbf{\Gamma}_l, \forall l,$ and $\mathbf{F}_l^{\star}, \forall l,$ are all positive semidefinite, problem (\ref{pr:digital 2}) can be solved by many toolboxes, e.g., CVX.
In the following, we provide a better method by deriving the closed-form solution of digital beamformer with lower computational complexity.
Based on Lagrange multiplier method, we first introduce $L$ Lagrangian multipliers $\beta_l \geq 0, \forall l,$ for each BS to substitute the transmit power constraints (\ref{pr:digital 2}b) into the objective function (\ref{pr:digital 2}a) and then the Lagrangian function can be given by
\begin{equation}
\label{fun lagrangian}
\begin{aligned}
\mathcal{\psi}(\mathbf{f}_{\textrm{BB}_{l,k}}) &= \sum_{l=1}^{L} \sum_{k=1}^{K} \omega_k\alpha_{l,k}\mathbf{f}_{\textrm{BB}_{l,k}}^H \mathbf{\Gamma}_l \mathbf{f}_{\textrm{BB}_{l,k}}- \sum_{l=1}^{L} \beta_l P_{\textrm{max}}\\~
&~-\sum_{l=1}^{L} \hspace{-0.2em}  \sum_{k=1}^{K} \hspace{-0.2em} 2 \hspace{-0.2em} \sqrt{\omega_k(1+\rho_k^{\star})} \Re\{ (\xi_k^{\star})^* \alpha_{l,k} \mathbf{h}_{l,k}^H  \mathbf{F}_{\textrm{RF}_l}^{\star} \mathbf{f}_{\textrm{BB}_{l,k}} \}\\~
&~+\sum_{l=1}^{L} \sum_{k=1}^{K} \beta_l \mathrm{Tr}\{\alpha_{l,k} \mathbf{f}_{\textrm{BB}_{l,k}}^H\mathbf{F}_l^{\star}\mathbf{f}_{\textrm{BB}_{l,k}} \}.
\end{aligned}
\end{equation}
\nid Function (\ref{fun lagrangian}) is a concave function of variables $\{\mathbf{f}_{\textrm{BB}_{l,k}}\}_{\forall l, \forall k}$ and the closed-form solution can be obtained by setting $\partial \mathcal{\psi}/ \partial \mathbf{f}_{\textrm{BB}_{l,k}}$ to zero
\begin{equation}\label{solution fbb}
\begin{aligned}
\mathbf{f}_{\textrm{BB}_{l,k}}^{\star}(\beta_l) = &\Big(\omega_k\alpha_{l,k} \left( \mathbf{\Gamma}_l   + \beta_l \mathbf{F}_l^{\star} \right) \Big)^{\dag} \times \\~
&\sqrt{\omega_k(1+ \rho_k^{\star} )} \xi_k^{\star} \alpha_{l,k}(\mathbf{F}_{\textrm{RF}_l}^{\star})^H \mathbf{h}_{l,k},
\end{aligned}
\end{equation}
where $(\cdot)^\dag$ is the matrix pseudoinverse.
Considering the power constraints of all BSs, the optimal $\beta_l$ satisfies
\vspace{ -0.10 cm}
\begin{equation}
\label{solution lamda}
\begin{aligned}
\hspace{-0.9 em}\sum_{k =1}^{K} \mathrm{Tr}\bigg(\alpha_{l,k} \big(\mathbf{f}_{\textrm{BB}_{l,k}}(\beta_l)\big)^H \mathbf{F}_l^{\star}  \mathbf{f}_{\textrm{BB}_{l,k}}(\beta_l) \bigg) - P_{\textrm{max}} = 0.
\end{aligned}
\end{equation}
We can optimize the optimal multipliers $\beta_l^{\star}$ by bisection search method.

In summary, the hybrid beamforming design algorithm for fully connected architecture is implemented as follows.
In each iteration, we first initialize the hybrid beamformer $\{\mathbf{F}_{\text{RF}_l}^0, \mathbf{f}_{\textrm{BB}_{l,k}}^0\}_{\forall l,\forall k}$.
Next, with the obtained closed-form solutions, we update auxiliary vectors $\bm{\rho}^{\star}$ and $\bm{\xi}^{\star}$ with (\ref{optimal rho}) and (\ref{optimal xi}), respectively.
Then, the analog beamformer $\{\mathbf{F}_{\text{RF}_l}^{\star}\}_{\forall l}$ can be calculated by proposed RCG-based analog beamformer design algorithm.
Last, the digital beamformers $\{\mathbf{f}_{\textrm{BB}_{l,k}}^{\star}\}_{\forall l,\forall k}$ are optimized by the closed-form solution in (\ref{solution fbb}).
The algorithm iterates until convergence and the proposed hybrid beamforming design algorithm for fully connected architecture is summarized in Algorithm 2.
Besides, the proposed hybrid beamforming design algorithm has great potential to extend into the jointly design of hybrid beamformer and combiner for cooperative mmWave MIMO systems.

\begin{algorithm}[t] \small% algorithm 2 for hybrid beamforming
  \caption{Hybrid Beamforming Design Algorithm for Fully Connected Architecture}
  \label{alg:Algorithm 2}
  \begin{algorithmic}[1]
    \REQUIRE $L, N_{\textrm{T}}, N_{\textrm{RF}}, K, \{\alpha_{l,k}, \mathbf{h}_{l,k}\}_{\forall l, \forall k}.$
    \ENSURE  $\{\mathbf{F}_{\textrm{RF}_l}^{\star}, \mathbf{f}_{\textrm{BB}_{l,k}}^{\star}\}_{\forall l,\forall k}$.
    \STATE Initialize $\{\mathbf{F}_{\text{RF}_l}^0, \mathbf{f}_{\textrm{BB}_{l,k}}^0\}_{\forall l,\forall k}$.
    \REPEAT
    \STATE Calculate $\bm{\rho}^*$ with (\ref{optimal rho});
    \STATE Calculate $\bm{\xi}^*$ with (\ref{optimal xi});
    \STATE Calculate $\mathbf{F}_{\textrm{RF}_l}^{\star}, \forall l,$ with Algorithm 1;
    \STATE Calculate $\mathbf{f}_{\textrm{BB}_{l,k}}^{\star}, \forall l,\forall k,$ with (\ref{solution fbb});
    \UNTIL convergence condition is met.
    \STATE Obtain $\mathbf{F}_{\textrm{RF}_l}^{\star}$ and $\mathbf{f}_{\textrm{BB}_{l,k}}^{\star}, \forall l,\forall k$.
  \end{algorithmic}
\vspace{ -0.00 cm}
\end{algorithm}

\subsection{Convergence and Computational Complexity Analysis}
In this subsection, we provide a brief analysis of computational complexity and convergence for the proposed hybrid beamforming design algorithm presented in Algorithm 2.

We start with analyzing the computational complexity.
In each iteration, the optimal solutions of $\bm{\rho}$ and $\bm{\xi}$ require approximately $\mathcal{O}(L K^2 N_{\textrm{T}} N_{\textrm{RF}})$ and $\mathcal{O}\big( L K (K+1) N_{\textrm{T}} N_{\textrm{RF}}\big)$ operations, respectively;
updating the analog beamformer $\{\mathbf{F}_{\text{RF}_l}\}_{\forall l}$ requires about $\mathcal{O}\big((L N_{\textrm{T}}N_{\textrm{RF}})^{1.5}\big)$ operations; the digital beamformer $\{\mathbf{f}_{\textrm{BB}_{l,k}}\}_{\forall l,\forall k}$ has a computational complexity of approximately $\mathcal{O}\big(I_a L K (N_{\textrm{T}}N_{\textrm{RF}}(K+N_{\textrm{RF}})+ N_{\textrm{RF}}^3)\big)$ and $I_a$ denotes the iterations of bisection search method.
Therefore, the total computational complexity of Algorithm 2 can be approximated by $\mathcal{O}\big(I_b ( I_a LK^2 N_{\textrm{T}} N_{\textrm{RF}} + (L N_{\textrm{T}}N_{\textrm{RF}})^{1.5} + I_a L K N_{\textrm{T}} N_{\textrm{RF}}^2+ I_a L K N_{\textrm{RF}}^3)  \big)$, wherein $I_b$ represents the required number of iterations for the algorithm convergence.

Then, we turn to demonstrate the convergence of algorithm.
Essentially, Algorithm 2 is a BCD algorithm for the reformulated problem (\ref{pr:fourvariables}) and there are four optimized variables needed to be updated.
In each iteration, the optimal solution of $\bm{\rho}$, $\bm{\xi}$, and $\mathbf{f}_{\textrm{BB}_{l,k}}$ can be easily proved to be monotonic.
For variable $\mathbf{F}_{\text{RF}_l}$, Algorithm 1 has been proved to be convergent in \cite{P.-A. Absil}.
Since the other operations are all monotonic, the convergency of Algorithm 2 can be proved directly, which is validated through Fig. \ref{fig:Sumrate_vs_iter} at Sec. VI-A.
Due to the space limitation, please refer to Appendix A in \cite{K. Shen2} for detailed proof.

As for the speed of convergence, the total computational complexity of Algorithm 2 is approximated by $\mathcal{O}\big(I_b ( I_a LK^2 N_{\textrm{T}} N_{\textrm{RF}} + (L N_{\textrm{T}}N_{\textrm{RF}})^{1.5} + I_a L K N_{\textrm{T}} N_{\textrm{RF}}^2+ I_a L K N_{\textrm{RF}}^3)  \big)$, wherein $I_a$ and $I_b$ represent the iterations of bisection search method and the required number of iterations for the algorithm convergence, respectively.
Hence, two factors mainly affect the speed of convergence, i.e, the solution accuracy of Algorithm 2 and the related updating factors, e.g., the Armijo backtracking line search step size $b$ in Algorithm 1.
Fortunately, the overall computational complexity of Algorithm 1 is approximately $\mathcal{O}\big((LN_{\textrm{T}} N_{\textrm{RF}})^{1.5}\big)$, which is an efficient approach.
Hence, the convergence speed of our proposed algorithm is acceptable with the reasonable solution accuracy.

\section{Hybrid Beamforming Design For SubArray Architecture}
In this section, we extend the proposed hybrid beamforming design algorithm into subarray architecture (i.e., fixed subarray in Sec. V-A and dynamic subarray in Sec. V-B) to explore the EE performance of the cooperative mmWave massive MIMO system.
Motivated by the properties of block diagonal matrix, we first propose a heuristic algorithm to design the analog beamformer for fixed subarray architecture.
Then, K-means based antenna design algorithm is proposed for the dynamic subarray architecture.

\subsection{Fixed SubArray Architecture}
Although the fully connected architecture has the potential of obtaining full beamforming gain for each RF chain, the required hardware cost and power consumption are still high for large-scale antenna arrays.
Thus, the fixed subarray architecture is introduced as an alternative architecture to implement the hybrid beamforming with reduced hardware cost and lower power consumption.
In this subsection, we extend the proposed hybrid beamforming design algorithm into the fixed subarray architecture with $N_{\textrm{T}}$ antennas and $N_{\textrm{RF}}$ RF chains.
In the fixed subarray architecture, each RF chain only connects to a non-overlapping fixed subarray and the number of antennas per subarray is fixed as $N_{\textrm{T}}^{\textrm{Sub}}=N_{\textrm{T}}/N_{\textrm{RF}}$.
Thus, the analog beamformer $\mathbf{F}_{\textrm{RF}_l}^{\textrm{fs}}$ of the $l$-th BS can be expressed as a block diagonal matrix
\bea \label{eq:frf for fs1}
\mathbf{F}_{\textrm{RF}_l}^{\textrm{fs}} = \left[ \begin{array}{cccc}
      \mathbf{f}_{\textrm{RF}_{l,1}}^{\textrm{fs}} & \mathbf{0} & \ldots & \mathbf{0} \\
    \mathbf{0} & \mathbf{f}_{\textrm{RF}_{l,2}}^{\textrm{fs}} &  & \mathbf{0} \\
        \vdots &   & \ddots & \vdots \\
        \mathbf{0} & \mathbf{0} & \ldots & \mathbf{f}^{\textrm{fs}}_{\textrm{RF}_{l,N_{\textrm{RF}}}}
                             \end{array}  \right], \forall l,
\eea
where $\mathbf{f}_{\textrm{RF}_{l,r}}^{\textrm{fs}} \in \mathbb{C}^{N_{\textrm{T}}^{\textrm{Sub}}}, l =1,\ldots, L, r =1,\ldots, N_{\textrm{RF}},$ is the analog beamformer vector associated with the $r$-th RF chain in the $l$-th BS.

In the following, we adopt the same idea proposed in Sec. IV-C to optimize the analog beamformer for fixed subarray architecture and the vectorization operation is the key step.
Benefit from the properties of block diagonal matrix, matrices $\mathbf{F}_{\textrm{RF}_l}^{\textrm{fs}}, \forall l,$ contain many zero elements, that is, elements that do not need to be optimized.
Thus, we first extract the non-zero elements from matrices $\mathbf{F}_{\textrm{RF}_l}^{\textrm{fs}}\in \mathbb{C}^{N_{\textrm{T}} \times N_{\textrm{RF}}}, \forall l,$ by defining $L$ auxiliary vectors $\mathbf{f}_l^{\textrm{fs}} \in \mathbb{C}^{N_{\textrm{T}}},\forall l$, as
\begin{equation}
\begin{aligned}\label{eq:f fs l}
\mathbf{f}_l^{\textrm{fs}} = \bigg[\big((\mathbf{f}_{\textrm{RF}_{l,1}}^{\textrm{fs}})^* \big)^T,\ldots,\big((\mathbf{f}^{\textrm{fs}}_{\textrm{RF}_{l,N_{\textrm{RF}}}})^*\big)^T\bigg]^T , \forall l.
\end{aligned}
\end{equation}
Compared with the variables in function $\tilde{f}_{a1}\big(\mathrm{vec}(\mathbf{F}_{\text{RF}_l})^*\big)$ (\ref{eq:varphi vec}), the auxiliary vector $\mathbf{f}_l^{\textrm{fs}}$ is actually part of $\mathrm{vec}(\mathbf{F}_{\text{RF}_l})^*, \forall l.$
Similarly, based on the reformulated function (\ref{eq:varphi vec}), we also choose useful information from the introduced variables $\tilde{\mathbf{t}}_{l},\forall l,$ and $\tilde{\mathbf{T}}_{l},\forall l,$ which can be given by
\begin{subequations}
\begin{align}
&\tilde{\mathbf{t}}_{l}^{\textrm{fs}}\triangleq \hat{\mathbf{E}}  \tilde{\mathbf{t}}_{l},\\
&\tilde{\mathbf{T}}_{l}^{\textrm{fs}} \triangleq \hat{\mathbf{E}}  \tilde{\mathbf{T}}_{l} \hat{\mathbf{E}}^H,
\end{align}
\end{subequations}
where $\hat{\mathbf{E}} \in \mathbb{R}^{N_{\textrm{T}} \times N_{\textrm{T}}N_{\textrm{RF}}}$ is the selection matrix.
The sub-matrices $\hat{\mathbf{E}}\big(:,(r\text{-}1)N_{\textrm{T}}\text{+}1:rN_{\textrm{T}}\big), r =1,\ldots, N_{\textrm{RF}},$ have all zeros elements except for the identity matrix $\mathbf{I}_{N_{\textrm{T}}^{\textrm{Sub}}}$ in the rows from $(r-1)N_{\textrm{T}}^{\textrm{Sub}}+1$ to $rN_{\textrm{T}}^{\textrm{Sub}},\forall r$.
Similar to Sec. IV-C, by introducing a transformation matrix $\bm{\Xi}_l$ for each BS and an intermediate vector $\mathbf{f}^{\textrm{fs}} \triangleq [(\mathbf{f}_1^{\textrm{fs}})^T, \ldots, (\mathbf{f}_{L}^{\textrm{fs}})^T]^T \in \mathbb{C}^{LN_{\textrm{T}}}$, we can rewrite the objective function of analog beamforming design for fixed subarray architecture as
\begin{equation}\label{eq:f fs 2}
\begin{aligned}
f_{s}(\mathbf{f}^{\textrm{fs}})= 2\Re \{ (\mathbf{f}^{\textrm{fs}})^H\mathbf{v}^{\textrm{fs}}\}- (\mathbf{f}^{\textrm{fs}})^H\mathbf{W}^{\textrm{fs}}\mathbf{f}^{\textrm{fs}},
\end{aligned}
\end{equation}
where two matrices are defined as $\mathbf{v}^{\textrm{fs}} \triangleq \sum_{l = 1}^{L} \mathbf{\Xi}_l\tilde{\mathbf{t}}_{l}^{\textrm{fs}}$ and $\mathbf{W}^{\textrm{fs}} \triangleq \sum_{l = 1}^{L}  \mathbf{\Xi}_l \tilde{\mathbf{T}}_{l }^{\textrm{fs}}\mathbf{\Xi}_l^H$ for brevity.
Hence, the sub-problem of analog beamformer design for fixed subarray architecture can be expressed as
\begin{subequations}\label{pr:f fs 3}
\begin{align}
\underset{ \mathbf{f}^{\textrm{fs}}} {\min} &~~-f_{s}(\mathbf{f}^{\textrm{fs}}) \\
\mathrm{s.t.}&~~| \mathbf{f}^{\textrm{fs}}(i)| = 1,~\forall i.
\end{align}
\end{subequations}
With RCG-based analog beamformer algorithm in Algorithm 1, the intermediate vector $\mathbf{f}^{\textrm{fs}}$ can be easily obtained.
Finally, the analog beamformer $\mathbf{F}_{\textrm{RF}_l}^{\textrm{fs}},\forall l,$ for fixed subarray architecture can be selected and reconstructed from $\mathbf{f}^{\textrm{fs}}$ through
\begin{subequations}
\begin{align}\label{analog for fixed}
&\mathbf{f}_{l}^{\textrm{fs}} = \bigg(\mathbf{f}^{\textrm{fs}}\big((l-1)N_{\textrm{T}}+1:lN_{\textrm{T}}\big)\bigg)^*,\\
&\mathbf{f}_{l,r}^{\textrm{fs}} = \mathbf{f}_{l}^{\textrm{fs}}\big((r-1)N_{\textrm{T}}^{\textrm{Sub}}+1:r N_{\textrm{T}}^{\textrm{Sub}} \big),\\
&\mathbf{F}_{\textrm{RF}_l}^{\textrm{fs}} = \mathrm{blkdiag} \{\mathbf{f}_{\textrm{RF}_{l,1}}^{\textrm{fs}}, \ldots, \mathbf{f}^{\textrm{fs}}_{\textrm{RF}_{l,N_{\textrm{RF}}}} \}.
\end{align}
\end{subequations}

\subsection{Dynamic SubArray Architecture}
\label{sec:PSA}
Unlike the fixed subarray architecture, the dynamic subarray architecture has higher degrees of freedom (DoF) on the hybrid beamforming design and can improve EE performance in mmWave massive MIMO systems.
In this section, we consider the hybrid beamforming design for dynamic subarray architecture.
Let $\mathcal{S}_{l,q}$ be the set of antenna elements, which are connected to the $q$-th RF chain of the $l$-th BS, $ q = 1, \ldots, N_{\textrm{RF}}$, $l=1,\ldots, L$.
The antenna elements in each set $\mathcal{S}_{l,q}$ will be dynamically designed to construct a subarray.

Taking hardware constraints of dynamic subarray architecture into account, when the introduced variables $\bm{\rho}$, $\bm{\xi}$, and digital beamformer $\{\mathbf{f}_{\textrm{BB}_{l,k}}\}_{\forall l,\forall k}$ are fixed, the sub-problem of analog beamformer for dynamic subarray architecture is
\begin{subequations}\label{pr:analog of dynamic1}
\begin{align}
\underset{ \{\mathbf{F}_{\textrm{RF}_l}^{\textrm{ds}}\}_{\forall l} } {\max}&\tilde{f}_{a}(\mathbf{F}_{\text{RF}_l}^{\textrm{ds}}) \\
\mathrm{s.t.}~~&\| \mathbf{F}_{\textrm{RF}_l}^{\textrm{ds}}(i,:) \|_0 = 1,~\forall l,i,\\
&| \mathbf{F}_{\textrm{RF}_l}^{\textrm{ds}}(i,q) | \in \{0, 1\},~\forall l,i,q,
\end{align}
\end{subequations}
where (\ref{pr:analog of dynamic1}b) indicates that each antenna element is connected to only one RF chain, (\ref{pr:analog of dynamic1}c) is the constant modulus constraint of PSs, and analog beamformer has the form of $\mathbf{F}_{\textrm{RF}_l}^{\textrm{ds}} \triangleq [ \mathbf{f}_{\textrm{RF}_{l,{\mathcal{S}_{l,1}}}}, \ldots, \mathbf{f}_{\textrm{RF}_{l,{\mathcal{S}_{l,N_{\textrm{RF}}}}}}]$.
To be specific, if the $i$-th antenna element is connected to the $q$-th RF chain of BS-$l$, i.e., the $i$-th antenna belongs to set $\mathcal{S}_{l,q}$, then $| \mathbf{F}_{\textrm{RF}_l}^{\textrm{ds}}(i,q) |=| \mathbf{f}_{\textrm{RF}_{l,{\mathcal{S}_{l,q}}}}(i)| = 1$; otherwise, $| \mathbf{F}_{\textrm{RF}_l}^{\textrm{ds}}(i,q) |=| \mathbf{f}_{\textrm{RF}_{l,{\mathcal{S}_{l,q}}}}(i)|=0$.
In order to indicate the matching between RF chains and antenna elements, we define $L$ binary matrices $\mathbf{V}_l$ for all BSs as
\begin{equation}
\label{eq:v of dynamic}
\begin{aligned}
\mathbf{V}_l = [ \mathbf{v}_{{\mathcal{S}_{l,1}}}, \ldots, \mathbf{v}_{{\mathcal{S}_{l,N_{\textrm{RF}}}}}],
\end{aligned}
\end{equation}
where $\mathbf{v}_{{\mathcal{S}_{l,q}}} \in \{0, 1\}^{N_{\textrm{T}}}$, $~ l =1, \ldots, L,~ q = 1, \ldots, N_{\textrm{RF}}$.
If the $i$-th antenna element belongs to set $\mathcal{S}_{l,q}$, $\mathbf{v}_{{\mathcal{S}_{l,q}}}(i) = 1$; otherwise, $\mathbf{v}_{l,{\mathcal{S}_{l,q}}}(i)  = 0$.
Then, the analog beamformer $\{\mathbf{F}_{\textrm{RF}_l}^{\textrm{ds}}\}_{\forall l}$ can be rewritten as:
\begin{equation}
\label{eq:beamforming of dynamic}
\begin{aligned}
\tilde{\mathbf{F}}_{\textrm{RF}_l}^{\textrm{ds}} = \mathbf{F}_{\textrm{RF}_l}^{\textrm{fc}} \odot \mathbf{V}_l.
\end{aligned}
\end{equation}
\nid Thus, problem (\ref{pr:analog of dynamic1}) can be transformed into
\begin{subequations}\label{pr:analog of dynamic2}
\begin{align}
\underset{ \{\mathbf{F}_{\textrm{RF}_l}^{\textrm{fc}} , \mathbf{V}_l \}_{\forall l} } {\max} & \tilde{f}_{a}(\tilde{\mathbf{F}}_{\textrm{RF}_l}^{\textrm{ds}}) \\
\mathrm{s.t.}~~~&\|\mathbf{V}_l(i,:) \|_0 = 1,~\forall l,i,\\
&| \mathbf{F}_{\textrm{RF}_l}^{\textrm{fc}}(i,q) | = 1,~\forall l,i,q.
\end{align}
\end{subequations}
However, problem (\ref{pr:analog of dynamic2}) is still an intractable problem due to the $l_0$-norm constraint on the binary matrices $\mathbf{V}_l,l =1, \ldots, L$.
In an effort to obtain a suitable solution, we propose a two-step sub-optimal solution to optimize problem (\ref{pr:analog of dynamic2}) by dividing the original problem into two sub-problems: \textit{i}) Analog beamforming design; \textit{ii}) Binary matrices calculation.

As for the analog beamforming design sub-problem, we can consider it as a similar problem in Sec. IV-C and then obtain the analog beamformer $\{\mathbf{F}_{\textrm{RF}_l}^{\textrm{fc}}\}_{\forall l}$ by the RCG-based analog beamforming design algorithm.
For the binary matrices calculation, we observe that the design of $\mathbf{V}_l,l =1, \ldots, L,$ is equivalent to optimizing the connection between RF chains and antenna elements, i.e., the antenna cluster sets $\{\mathcal{S}_{l,q}\}_{q = 1}^{N_{\textrm{RF}}}$ for the $l$-th BS.
This observation motivates us to transform the sub-problem of $\mathbf{V}_l,l =1, \ldots, L,$ into an antenna grouping problem, which optimizes the antenna cluster sets $\{\mathcal{S}_{l,q}\}_{q = 1}^{N_{\textrm{RF}}}$ for all BSs.

% 算法
\begin{algorithm}[t]\small
  \caption{K-means based Antenna Design Algorithm}
  \label{alg:Algorithm 3}
  \begin{algorithmic}[1]
    \REQUIRE $\{\mathbf{F}_{\textrm{RF}_l}^{\textrm{fc}}\}_{\forall l}, \mathcal{S}_{0}= \{ 1,\ldots, N_{\textrm{T}}\}, Iter$.
    \ENSURE  $\{\mathcal{S}_{l,q}\}_{q=1}^{N_{\textrm{RF}}},\forall l$.
    \FOR{$l = \{1 : L\}$}
    \STATE Calculate equivalent metric $\mathbf{R}_{l} = \mathbf{F}_{\textrm{RF}_l}^{\textrm{fc}}(\mathbf{F}_{\textrm{RF}_l}^{\textrm{fc}})^H$.
    \STATE Choose $N_{\textrm{RF}}$ initial cluster centers randomly and define centroid set as $\mathcal{S}_C$, initialize $iter = 1$.
    \REPEAT
    \STATE Group each antenna element into group $\{\mathcal{S}_{j}\}_{j=1}^{N_{\textrm{RF}}}$.
    \STATE Batch update:
    \FOR{$i = \{\mathcal{N}_{t}-\mathcal{S}_C\}$}
    \STATE $j^*$ = $\mathrm{arg}~  \underset{j \in \mathcal{S}_C}{\mathrm{max}} ~ \hat{\zeta}(i,\mathcal{S}_{j})$;
     \STATE  $\mathcal{S}_{j^*} \leftarrow  \mathcal{S}_{j^*} \cup i$;
     \STATE  $\mathcal{S}_{0} \leftarrow \mathcal{S}_{0} \setminus i$.
    \ENDFOR
    \STATE Compute average mutual information of each cluster to obtain $N_{\textrm{RF}}$ new cluster centers and update $\mathcal{S}_C$.
    \STATE $iter = iter + 1$
    \UNTIL $\{\mathcal{S}_{j}\}_{j=1}^{N_{\textrm{RF}}}$ does not change or $iter >= Iter$.
    \STATE Update $\{\mathcal{S}_{l,q}\}_{q=1}^{N_{\textrm{RF}}} := \{\mathcal{S}_{j}\}_{j=1}^{N_{\textrm{RF}}}$.
    \ENDFOR
  \end{algorithmic}
\end{algorithm}

Then, in order to group antennas into different clusters, we need to define a performance metric that distinguishes the correlation between antennas.
The stronger correlation, the more likely the antenna elements are to be assigned to the same antenna group.
Fortunately, the Minkowski distance (i.e., $\|\mathbf{x}_i - \mathbf{x}_j\|_p$) is the definition of a group of distances and widely used for dynamic subarray design in literatures, such as \cite{S. Park}, \cite{Y. Sun}.
\nid Specifically, when $p$ = $2$, the Minkowski distance is equivalent to the Euclidean distance.
Actually, all of these variants of Minkowski distance can be selected as the performance metric for antenna grouping algorithm.
In order to ensure the fairness of algorithm comparison, we use Minkowski $l_2$-norm \cite{Y. Sun} as a new equivalent metric to solve the grouping problem, which can be expressed as
\begin{equation}
\label{eq:equivalent metric}
\begin{aligned}
\zeta(\mathbf{R}_{\mathcal{S}_{l,q}}) = \frac{1}{|\mathcal{S}_{l,q}|} \sum\limits_{i \in \mathcal{S}_{l,q}} \sum\limits_{j \in \mathcal{S}_{l,q}} |\mathbf{R}_{\mathcal{S}_{l,q}}(i,j)|,
\end{aligned}
\end{equation}
where $\mathbf{R}_{\mathcal{S}_{l,q}} \triangleq \mathbf{f}_{\textrm{RF}_{l,{\mathcal{S}_{l,q}}}}(\mathbf{f}_{\textrm{RF}_{l,{\mathcal{S}_{l,q}}}})^H \in \mathbb{C}^{N_{\textrm{T}} \times N_{\mathrm{T}}}$.
The Minkowski $l_2$-norm can be interpreted as the correlation between antenna elements in set $\mathcal{S}_{l,q}$.
The stronger correlation, the more likely the antenna elements are to be assigned to the same antenna group.
Similarly, we also define the correlation between the different groups as
\begin{equation}
\label{eq:equivalent metric2}
\begin{aligned}
\hat\zeta(\mathcal{S}_{l,i},\mathcal{S}_{l,j}) = \frac{1}{|\mathcal{S}_{l,i}||\mathcal{S}_{l,j}|} \hspace{-0.2 em} \sum\limits_{m \in \mathcal{S}_{l,i}} \hspace{-0.2 em} \sum\limits_{n \in \mathcal{S}_{l,j}}  \hspace{-0.2 em} |\mathbf{R}_{l}(m,n)|,
\end{aligned}
\end{equation}
where $\mathbf{R}_{l} \triangleq \mathbf{F}_{\textrm{RF}_l}^{\textrm{fc}}(\mathbf{F}_{\textrm{RF}_l}^{\textrm{fc}})^H ,l =1, \ldots, L $.
Hence, the dynamic subarray sub-problem of antenna grouping is reformulated as
\begin{subequations}
\label{pr:dynamic of clustering}
\begin{align}
\underset{\{ \mathcal{S}_{l,1}, \cdots, \mathcal{S}_{l,N_{\textrm{RF}}} \}_{\forall l} } {\max} & \sum_{l=1}^{L} \sum_{q=1}^{N_{\textrm{RF}}} \zeta(\mathbf{R}_{\mathcal{S}_{l,q}})\\
\mathrm{s.t.}~~~~~&\cup_{q=1}^{N_{\textrm{RF}}}\mathcal{S}_{l,q} = \{1, \ldots, N_{\textrm{T}}\}, \forall l,\\
&\mathcal{S}_{l,i} \cap \mathcal{S}_{l,j} = \varnothing, \forall l,i \neq j,\\
&\mathcal{S}_{l,q} \neq \varnothing, \forall l,q.
\end{align}
\end{subequations}

\nid The optimization of (\ref{pr:dynamic of clustering}) is a combinatorial problem, which can be solved by exhaustive search algorithm with high computational complexity.
The large-scale antenna arrays, often used in mmWave massive MIMO systems, will make the exhaustive search algorithm unaffordable.
To reduce the computational complexity, we propose a low-complexity K-means based antenna design algorithm.
Benefit from the K-means clustering algorithm in \cite{T. Kanungo}, problem (\ref{pr:dynamic of clustering}) can be deemed as a classification problem, which needs to classify $N_{\textrm{T}}$ antennas to $N_{\textrm{RF}}$ groups for each BS.
Thus, based on the formula (\ref{eq:equivalent metric}) and (\ref{eq:equivalent metric2}), we propose K-means based antenna design algorithm to cluster the antennas into different subarrays connected to the corresponding RF chains.
The whole antenna grouping process is summarized in Algorithm 3.
In the following, we provide a brief computational complexity analysis for the proposed K-means based antenna design algorithm.
As shown in Algorithm 3, the calculation of equivalent metric $\mathbf{R}_l$ requires $\mathcal{O}(N_{\textrm{T}}^2 N_{\textrm{RF}})$ operations; and then the computational complexity of grouping the antennas is approximately at most $\mathcal{O}(I_c N_{\textrm{T}} N_{\textrm{RF}})$, wherein the parameter $I_c$ is the iterations of grouping each antenna element into clusters.
Meanwhile, the antenna elements of all $L$ BSs need to be grouped.
Hence, the total computational complexity of Algorithm 3 is approximated by $\mathcal{O}(L N_{\textrm{T}}^2 N_{\textrm{RF}})$.
Besides, it is of great interest to explore more advanced K-means algorithms, such as K-means++, to tackle the antenna grouping problem in future studies.

After obtaining the antenna groups $\mathcal{S}_{l,q}, ~\forall l, \forall q$, the binary matrices $\mathbf{V}_l, ~ \forall l,$ is obtained by (\ref{eq:v of dynamic}) and then the sub-optimal solution of analog beamformer $\{\tilde{\mathbf{F}}_{\textrm{RF}_l}^{\textrm{ds}}\}_{\forall l}$ can be updated by (\ref{eq:beamforming of dynamic}).
Finally, the digital beamformer $\{\mathbf{f}_{\textrm{BB}_{l,k}}\}_{\forall l, \forall k}$ are calculated by the same algorithm proposed in Sec. IV-D.

\section{Simulation Results}
\label{sc:Simulation}
%仿真参数说明
In this section, the achievable weighted sum-rate performance and EE performance of proposed algorithms for three typical antenna array architectures are provided through many extensive simulation results.
We consider that $L=3$ BSs and $K=9$ single-antenna users are distributed over an area randomly.
$N_{\textrm{T}}=48$ antennas and $N_{\textrm{RF}}=3$ RF chains are connected to each BS.
We assume that all users have the same priority (i.e., $\omega_k = 1, \forall k$) to assure the fairness among $K$ users.
For all simulations, the complex gain $\upsilon_n$ is generated by a complex Gaussian distribution $\upsilon_n \sim \mathcal{CN}(0, 10^{-0.1\kappa})$.
The other parameters are setting as follows: $\kappa = \kappa_a + 10\kappa_b \mathrm{log}10(d) + \kappa_c$, $\kappa_c \sim \mathcal{CN}(0, \sigma_c^2)$, $\kappa_a = 32$, $\kappa_b = 2$, $\sigma_c^2 = 8.7 $dB, and the variance of noise $\sigma_k^2, \forall k$, is $-20$ dBm.
All simulation results are based on over $10^4$ channel generations.

\subsection{Convergence behavior}
\begin{figure}[t]
\centering
  \includegraphics[width= 3.0 in]{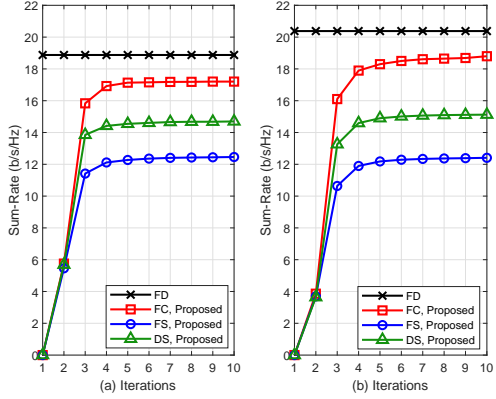}
  \vspace{ -0.1 cm}
  \caption{Achievable sum-rate versus number of iterations (a): $L=3, K=9$, $N_{\textrm{RF}}=3$, $N_{\textrm{T}}=48$, $P_{\textrm{max}}= 20$ dB; (b): $L=4, K=8$, $N_{\textrm{RF}}=2$, $N_{\textrm{T}}=32$, $P_{\textrm{max}}= 20$ dB.}
  \vspace{ -0.0 cm}
  \label{fig:Sumrate_vs_iter}
\end{figure}

%收敛性曲线-分析
In Fig. \ref{fig:Sumrate_vs_iter}, we first show the convergence of proposed algorithms with different system scenarios by plotting the weighted achievable sum-rate versus the number of iterations, where the transmit power is fixed at $P_{\textrm{max}}$ = 20 dB.
For the sake of illustration, we label the fully connected architecture as ``FC", the fixed subarray architecture as ``FS", and the dynamic subarray architecture as ``DS", respectively.
Simulation results in Fig. \ref{fig:Sumrate_vs_iter} indicate that the convergence of proposed algorithms is fast.
The weighted achievable sum-rate of proposed algorithms gradually increases with the increasing number of iterations and converges to a stationary point within about seven iterations.

%收敛性曲线-图

\subsection{Sum-Rate Performance}
Next, we compare the weighted achievable sum-rate performance of different hybrid beamforming optimization algorithms for various system parameters.
%对比算法说明
In order to better verify the effectiveness of proposed algorithms, we consider the following algorithms for comparison:
\begin{itemize}
\item \textit{FD}: Fully digital beamformer is obtained by the method proposed in Sec. IV-D.
\item \textit{FC, SDR} \cite{Q. Hou}: In this case, each RF chain connects all antenna elements with a fully connected architecture. Jointly design the user association scheme and analog beamformer with pre-defined codebook which has $N_{\mathrm{c}} = 64$ beamforming vectors that can be selected. Then, the digital beamforming vectors are solved by the SDR technique.
\item \textit{FS, SDR} \cite{Q. Hou}: Similar to the ``FC, SDR" algorithm, this algorithm considers the fixed subarray architecture and performs the hybrid beamforming with pre-defined codebook and SDR technique.
\item \textit{DS, S-AHC} \cite{Y. Sun}: Considering hybrid beamforming design with the dynamic subarray architecture, the analog beamformer is designed by shared agglomerative hierarchical clustering (S-AHC) algorithm in \cite{Y. Sun}.
\end{itemize}

%dB Vs Sum-Rate曲线图
\begin{figure}[t]
\centering
  \includegraphics[width = 3.0 in]{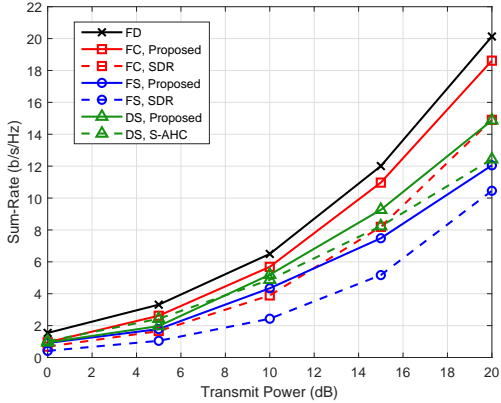}
  \vspace{ -0.1 cm}
  \caption{Achievable sum-rate versus transmit power ($L=3, K=9$, $N_{\textrm{RF}}=3$, $N_{\textrm{T}}=48$).}
  \vspace{ -0.0 cm}
  \label{fig:Sumrate_vs_snr}
\end{figure}

%PdB Vs Sum-Rate 分析
In Fig. \ref{fig:Sumrate_vs_snr}, we plot the sum-rate performance versus transmit power $P_{\mathrm{max}}$.
We can observe that the proposed algorithm with fully connected architecture has a satisfactory performance that close to the ``FD" scheme with the increasing transmit power.
It also shows the great advantages of fully connected hybrid beamforming architecture than other architectures.
Furthermore, the proposed algorithm for the dynamic subarray architecture and fixed subarray architecture outperform the ``DS, S-AHC " and ``FS, SDR" algorithms, respectively.
The results in Fig. \ref{fig:Sumrate_vs_snr} further indicate that the dynamic subarray hybrid beamforming architecture can obtain satisfactory performance while reducing hardware complexity.

%Antennas Vs Sum-Rate 曲线图
\begin{figure}[t]
\centering
  \includegraphics[width= 3.0 in]{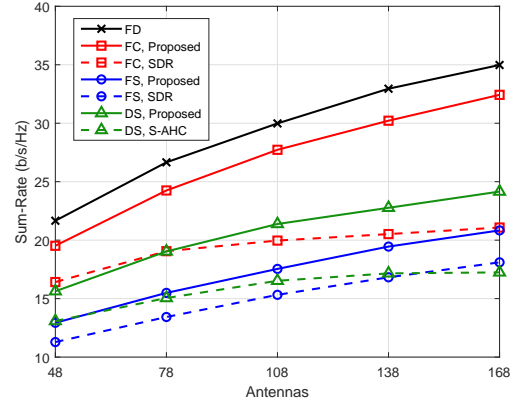}
  \vspace{ -0.1 cm}
  \caption{Achievable sum-rate versus number of antennas ($L=3, K=9$, $N_{\textrm{RF}}=3$, $P_{\textrm{max}}= 20$ dB).}
  \vspace{ -0.0 cm}
  \label{fig:Antennas_Vs_Sumrate}
\end{figure}

%Antennas Vs Sum-Rate 分析
Fig. \ref{fig:Antennas_Vs_Sumrate} indicates the sum-rate performance as a function of the number of antennas $N_{\textrm{T}}$.
In this case, we fix the transmit power at $P_{\textrm{max}}= 20$ dB.
The results in Fig. \ref{fig:Antennas_Vs_Sumrate} show that the achievable sum-rate increases with the increasing number of transmit antennas and the advantages of proposed algorithms are gradually prominent with the increasing of antennas compared with other algorithms.
When the number of antennas is more than $N_{\textrm{T}} = 78$, the performance of ``DS, Proposed" algorithm is even better than that of ``FC, SDR" algorithm, which demonstrates the advantages of dynamic subarray hybrid beamforming architecture with large-scale antenna arrays.

%RF chain Vs Sum-Rate 曲线图
\begin{figure}[t]
\centering
  \includegraphics[width= 3.0 in]{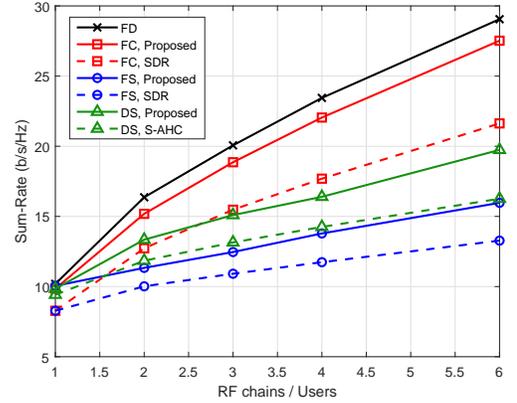}
  \vspace{ -0.0 cm}
  \caption{Achievable sum-rate versus number of RF chains / Users ($L=3$, $N_{\textrm{T}}=48$, $P_{\textrm{max}}= 20$ dB).}
  \vspace{ -0.05 cm}
  \label{fig:RFchain_Vs_Sumrate}
\end{figure}

\begin{table}[t]
\centering
\caption{Hardware Devices Power Consumption Parameters}     %标题
\begin{center}\label{tab:power consumption}
\begin{tabular}{cc} %两列，居左，中，左
   \toprule    %第一条线
   Hardware Devices    & Power consumption (mW) \\
   \midrule    %第二条线
   Baseband Processor & $P_{\textrm{BB}}$ = 200 \cite{R. Mendez-Rial}  \\
   RF Chain           & $P_{\textrm{RF}}$ = 300 \cite{R. Mendez-Rial}  \\
   Phase Shifter      & $P_{\textrm{PS}}$ = 25  \cite{R. Mendez-Rial} \\
   Switch             & $P_{\textrm{SW}}$ = 5   \cite{H. Li}\\
   \bottomrule %第三条线
\end{tabular}
\end{center}
\end{table}

%RF chain Vs Sum-Rate 分析
Fig. \ref{fig:RFchain_Vs_Sumrate} illustrates the sum-rate performance against the number of RF chains/users $N_{\textrm{RF}}/K$, which can be regarded as the sum-rate performance evaluation in denser scenarios.
In this case, the number of users served by each BS is equal to the number of RF chains and the number of antennas is fixed at $N_{\textrm{T}}=48$.
Fig. \ref{fig:RFchain_Vs_Sumrate} further confirms the performance of proposed user association and hybrid beamforming algorithm with different RF chains/users and the proposed algorithm can still achieve satisfactory performance even in dense scenarios.
With larger number of RF chains/users, each user is served by fewer antennas so that the performance gap between fully connected architecture and subarray architecture becomes larger, which illustrates the fact that when the number of antennas $N_{\textrm{T}}$ per BS is fixed, the performance of the proposed algorithm with dynamic subarray architecture will be limited as the number of RF chains/users increases.

\subsection{EE Performance}
In order to verify the trade-off between sum-rate performance and energy consumption with various hybrid beamforming architectures, we evaluate the EE performance of proposed algorithms, which is defined as \cite{X. Yu}
\begin{equation}
\label{eq:ee fun}
\begin{aligned}
\eta = \frac{R_{\rm{sum}}}{P_{\textrm{tot}}},
\end{aligned}
\end{equation}
where the unit of $\eta$ is b/s/Hz/W, $R_{\rm{sum}}$ and $P_{\textrm{tot}}$ are the sum of weighted achievable rate of all users and the total power consumption of all BSs, respectively.
Then, the total power consumption $P_{\textrm{tot}}$ of various antenna architectures can be commonly defined as
\begin{equation}
\label{eq:ptot}
\begin{aligned}
P_{\textrm{tot}} = L(P_{\textrm{max}}+P_{\textrm{BB}}+P_{\textrm{HW}}),
\end{aligned}
\end{equation}
where $L$ is the number of BSs, $P_{\textrm{max}}$ is the maximum transmit power for each BS, $P_{\textrm{BB}}$ and $P_{\textrm{HW}}$ are the powers consumed by the baseband processor and the hardware circuits (i.e., RF chains, phase shifters, and switches), respectively.
For different antenna architectures, the hardware capacities power consumption $P_{\textrm{HW}}$ are given by
\begin{subequations}
\label{eq:ptot fd}
\begin{align}
P_{\textrm{HW}}^{\textrm{FD}} &= N_{\textrm{T}}P_{\textrm{RF}}, \\
P_{\textrm{HW}}^{\textrm{FC}} &= N_{\textrm{RF}}P_{\textrm{RF}}+N_{\textrm{T}}N_{\textrm{RF}}P_{\textrm{PS}},\\
P_{\textrm{HW}}^{\textrm{FS}} &= N_{\textrm{RF}}P_{\textrm{RF}}+N_{\textrm{T}}P_{\textrm{PS}}, \\
P_{\textrm{HW}}^{\textrm{DS}} &= N_{\textrm{RF}}P_{\textrm{RF}}+N_{\textrm{T}}P_{\textrm{PS}}+N_{\textrm{T}}P_{\textrm{SW}},
\end{align}
\end{subequations}
where $P_{\textrm{PS}}$ and $P_{\textrm{SW}}$ are the energy consumed by a phase shifter and a switch, respectively.
In this simulation subsection, the setting of power parameters in practical mmWave massive MIMO systems are given in Table \ref{tab:power consumption}.

%EE Vs PdB 曲线图
\begin{figure}[!t]
\centering
  \includegraphics[width = 3.0 in]{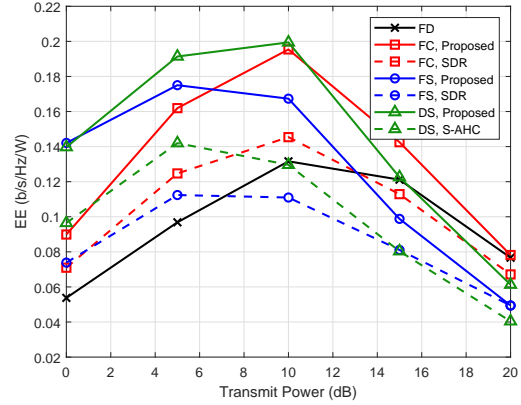}
  \vspace{-0.0 cm}
  \caption{EE versus transmit power ($L=3, K=9$, $N_{\textrm{RF}}=3$, $N_{\textrm{T}}=48$).}
  \vspace{-0.0 cm}
  \label{fig:EE_Vs_Sumrate}
\end{figure}

%EE Vs PdB  分析
In Fig. \ref{fig:EE_Vs_Sumrate}, we show the EE performance as a function of maximum transmit power $P_{\textrm{max}}$.
Due to the logarithmic function of sum-rate and the potential larger influence of $P_{\textrm{max}}$, the EE performance increases first and then decreases with the increase of $P_{\textrm{max}}$.
Although the ``FD'' scheme can achieve better sum-rate performance, the EE performance is poor due to the use of a large number of high power consumption RF chains.
In comparison, the proposed algorithm with the fully connected architecture has better EE performance.
Owing to the fewer number of phase shifters, the ``DS, Proposed" algorithm has the highest EE performance when the transmit power $P_{\textrm{max}}$ is less than 11.5 dB.

%EE Vs Nt 曲线图
\begin{figure}[t]
\centering
  \includegraphics[width = 3.0 in]{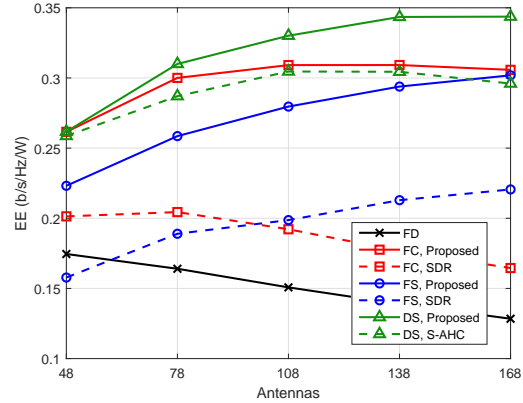}
  \vspace{-0.0 cm}
  \caption{EE versus number of antennas ($L=3, K=9$, $N_{\textrm{RF}}=3$, $P_{\textrm{max}}= 10$ dB).}
  \vspace{-0.0 cm}
  \label{fig:EE_Vs_Nt}
\end{figure}

%EE Vs Nt  分析
Next, Fig. \ref{fig:EE_Vs_Nt} demonstrates the EE performance with respect to the number of antennas $N_{\mathrm{T}}$.
With the increasing number of antennas, the ``FD'' scheme requires a larger number of RF chains, which causes much higher power consumption than other architectures.
Therefore, the EE performance of the ``FD'' scheme is worst and has a decreasing trend with the increasing number of antennas.
Moreover, the fully connected hybrid beamforming architecture enjoys a higher EE performance with fewer antennas.
Owing to the use of fewer phase shifters and relatively energy-saving switches, the fixed subarray and dynamic subarray architectures are more energy efficient with the large-scale antenna arrays.

%EE Vs RF chian 曲线图
\begin{figure}[t]
\centering
  \includegraphics[width = 3.0 in]{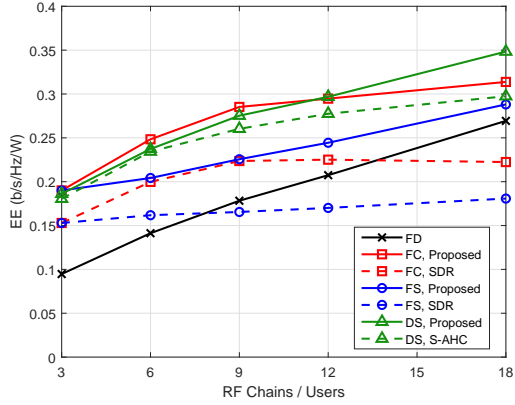}
  \vspace{-0.0 cm}
  \caption{EE versus number of RF chains / Users ($L=3$, $N_{\textrm{T}}=48$, $P_{\textrm{max}}= 10$ dB).}
  \vspace{-0.0 cm}
  \label{fig:EE_Vs_RFC}
\end{figure}

%EE Vs Nt  分析
Finally, Fig. \ref{fig:EE_Vs_RFC} represents the EE performance versus the number of RF chains and users.
The similar conclusion can be drawn from Fig. \ref{fig:EE_Vs_RFC} that the dynamic subarray hybrid beamforming architecture is vital and meaningful to reduce energy consumption for mmWave MIMO communications.
The ``FC, Proposed'' scheme maintains a higher EE performance when the RF chains/users is less than 12, since fewer RF chains are required but with high sum-rate performance.

\section{Conclusions}
In this paper, we proposed effective algorithms to solve the design of user association and hybrid beamforming for cooperative mmWave MIMO systems.
Aiming at maximizing the weighted sum-rate performance of the considered system, we divided the original problem into user association design and hybrid beamforming design sub-problems.
To be specific, the user association was designed by an equivalent problem and then an alternating maximization algorithm was proposed to iteratively optimize the analog and digital beamformer.
In order to reduce the hardware complexity, the algorithms were extended to design the hybrid beamformer with fixed subarray and dynamic subarray architectures, respectively.
The extensive simulation results have verified and confirmed the effectiveness of our proposed hybrid beamforming design algorithms in terms of both sum-rate performance and EE performance.
In our future work, more complex and comprehensive constraints (e.g., quality of service constraints), less signaling overhead algorithms (e.g., distributed beamforming), more innovative architectures (e.g., cell-free networks), and more complete designing (e.g., jointly design of hybrid beamformer and combiner) will be studied as the extended research directions.


\begin{thebibliography}{99}


\bibitem{glo2021}  P. Ni, R. Liu, M. Li, and Q. Liu, ``Hybrid analog-digital beamforming in cooperative mmWave MIMO systems," in \textit{Proc. IEEE Global Commun. Conf. (GLOBECOM)}, Madrid, Spain, Dec. 2021.

%%%% Overview of mmWave & channel
% mmWave
\bibitem{R. W. Heath}    R. W. Heath, N. G.-Prelcic, S. Rangan, W. Roh, and A. M. Sayeed, ``An overview of signal processing techniques for millimeter wave MIMO systems," \textit{IEEE J. Sel. Topics Signal Process.}, vol. 10, no. 3, pp. 436-453, Apr. 2016.

% hybrid beamforming proposed
\bibitem{X. Zhang2005}   X. Zhang, A. F. Molisch, and S.-Y. Kung, ``Variable-phase-shift-based RF-baseband codesign for MIMO antenna selection,"  \textit{IEEE Trans. Signal Process.}, vol. 53, no. 11, pp. 4091-4103, Nov. 2005.

%%%% three architectures
% fully connected
% mmWave + hybrid beamforming
\bibitem{T. Lin}         T. Lin, J. Cong, Y. Zhu, J. Zhang, and K. B. Letaief, ``Hybrid beamforming for millimeter wave systems using the MMSE criterion," \textit{IEEE Trans. Commun.}, vol. 67, no. 5, pp. 3693-3708, May 2019.

% ee performance
\bibitem{Y. Zhang}       Y. Zhang, \textit{et al.}, ``Energy-efficient hybrid precoding design for mm-Wave massive antenna multi-user systems," in \textit{Proc. Int. Symp. Antennas, Propagat. and EM Theory (ISAPE)}, Hangzhou, China, Dec. 2018.

\bibitem{C. Fang}        C. Fang, B. Makki, J. Li, and T. Svensson, ``Coordinated hybrid precoding for energy-efficient millimeter wave systems," in \textit{Proc. IEEE Int. Workshop Signal Process. Advances Wireless Commun. (SPAWC)}, Kalamata, Greece, Aug. 2018.

% fixed sub-array
\bibitem{X. Yu}          X. Yu, J. Shen, J. Zhang, and K. B. Letaief, ``Alternating minimization algorithms for hybrid precoding in millimeter wave MIMO systems," \textit{IEEE J. Sel. Topics Signal Process.}, vol. 10, no. 3, pp. 485-500, Apr. 2016.

\bibitem{J. Mo}         J. Mo, A. Alkhateeb, S. A.-Surra, and R. W. Heath, ``Hybrid architectures with few-bit ADC receivers: Achievable rates and energy-rate tradeoffs," \textit{IEEE Trans. Wireless Commun.}, vol. 16, no. 4, pp. 2274-2287, Apr. 2017.

\bibitem{X. Yu2}        X. Yu, F. Xu, K. Yu, and X. Dang, ``Power allocation for energy efficiency optimization in multi-user mmWave-NOMA system with hybrid precoding," \textit{IEEE Access}, vol. 7, pp. 109083-109093, Aug. 2019.

\bibitem{X. Zhao}       X. Zhao, T. Lin, Y. Zhu, and J. Zhang, ``Partially-connected hybrid beamforming for spectral efficiency maximization via a weighted MMSE equivalence," \textit{IEEE Trans. Wireless Commun.}, vol. 20, no. 12, pp. 8218-8232, Dec. 2021.

\bibitem{C. Fang2}      C. Fang, B. Makki, J. Li, and T. Svensson, ``Hybrid precoding in cooperative millimeter wave networks," \textit{IEEE Trans. Wireless Commun.}, vol. 20, no. 8, pp. 5373-5388, Aug. 2021.

% dynamic sub-array
\bibitem{S. Park}      S. Park, A. Alkhateeb, and R. W. Heath, ``Dynamic subarrays for hybrid precoding in wideband mmWave MIMO Systems," \textit{IEEE Trans. Wireless Commun.}, vol. 16, no. 5, pp. 2907-2920, May 2017.

\bibitem{Y. Sun}       Y. Sun, Z. Gao, H. Wang, and D. Wu, ``Machine learning based hybrid precoding for mmWave MIMO-OFDM with dynamic subarray," in \textit{Proc. IEEE Globecom Workshops (GC Wkshps)}, Abu Dhabi, United Arab Emirates, Dec. 2018.

\bibitem{K. Xu}        K. Xu, F. Zheng, P. Cao, H. Xu, and X. Zhu, ``A low complexity greedy algorithm for dynamic subarrays in mmWave MIMO systems," in \textit{Proc. IEEE Veh. Technol. Conf. (VTC2019-Fall)}, Honolulu, USA, Sep. 2019.

\bibitem{F. Yang}      F. Yang, \textit{et al.}, ``A partially dynamic subarrays structure for wideband mmWave MIMO systems," \textit{IEEE Trans. Commun.}, vol. 68, no. 12, pp. 7578-7592, Dec. 2020.

\bibitem{H. Li}        H. Li, M. Li, and Q. Liu, ``Hybrid beamforming with dynamic subarrays and low-resolution PSs for mmWave MU-MISO systems," \textit{IEEE Trans. Commun.}, vol. 68, no. 1, pp. 602-614, Jan. 2020.

\bibitem{H. Li JSAC}   H. Li, M. Li, Q. Liu, and A. L. Swindlehurst, ``Dynamic hybrid beamforming with low-resolution PSs for wideband mmWave MIMO-OFDM systems,'' \textit{IEEE J. Sel. Area Commun.}, vol. 38, no. 9, pp. 2168-2181, Sep. 2020.

\bibitem{L. Yan}       L. Yan, C. Han, and J. Yuan, ``A dynamic array-of-subarrays architecture and hybrid precoding algorithms for Terahertz wireless communications," \textit{IEEE J. Sel. Areas Commun.}, vol. 38, no. 9, pp. 2041-2056, Sep. 2020.

% Cooperative Networks
\bibitem{D. Gesbert2010}   D. Gesbert, S. Hanly, H. Huang, S. S. Shitz, O. Simeone, and W. Yu, ``Multi-cell MIMO cooperative networks: A new look at interference," \textit{IEEE J. Sel. Areas Commun.}, vol. 28, no. 9, pp. 1380-1408, Dec. 2010.

\bibitem{J. Zhang2019}     J. Zhang, S. Chen, Y. Lin, J. Zheng, B. Ai, and L. Hanzo, ``Cell-free massive MIMO: A new next-generation paradigm," \textit{IEEE Access}, vol. 7, pp. 99878-99888, Jul. 2019.

\bibitem{Z. Chen2019}      Z. Chen, F. Sohrabi, and W. Yu, ``Multi-cell sparse activity detection for massive random access: Massive MIMO versus cooperative MIMO," \textit{IEEE Trans. Wireless Commun.}, vol. 18, no. 8, pp. 4060-4074, Aug. 2019.

\bibitem{B. He2019}     B. He, Q. Ni, J. Chen, L. Yang, and L. Lv, ``User-pair selection in multiuser cooperative networks with an untrusted relay," \textit{IEEE Trans. Veh. Technol.}, vol. 68, no. 1, pp. 869-882, Jan. 2019.

% cooperative network + hybrid beamforming
\bibitem{I. Ahmed}     I. Ahmed, \textit{et al.}, ``A survey on hybrid beamforming techniques in 5G: Architecture and system model perspectives," \textit{IEEE Commun. Surveys Tutorials}, vol. 20, no. 4, pp. 3060-3097, Jun. 2018.

\bibitem{K. Li}        K. Li, Y. Yang, Y. Chen, X. Yang, and H. Yi, ``A novel network optimization method for cooperative massive MIMO systems," in \textit{Proc. IEEE Veh. Technol. Conf. (VTC-Spring)}, Sydney, Australia, Jun. 2017.

% user association
\bibitem{S. Tong}      S. Tong, Y. Liu, M. Cheriet, M. Kadoch, and B. Shen, ``UCAA: User-centric user association and resource allocation in fog computing networks," \textit{IEEE Access}, vol. 8, pp. 10671-10685, Jan. 2020.

% coopeartive network  + mmWave + hybrid beamforming
\bibitem{C. Lee}       C. Lee and W. Chung, ``Hybrid RF-baseband precoding for cooperative multiuser massive MIMO systems with limited RF chains," \textit{IEEE Trans. Commun.}, vol. 65, no. 4, pp. 1575-1589, Apr. 2017.

\bibitem{J. Jiang}      J. Jiang, Y. Yuan, and L. Zhen, ``Multi-user hybrid precoding for dynamic subarrays in mmWave massive MIMO systems," \textit{IEEE Access}, vol. 7, pp. 101718-101728, Jul. 2019.

% reference  35
\bibitem{Q. Hou}       Q. Hou, S. He, Y. Huang, H. Wang, and L. Yang, ``Joint user scheduling and hybrid precoding design for MIMO C-RAN," in \textit{Proc. Int. Conf. Wireless Commun. Signal Process. (WCSP)}, Nanjing, China, Oct. 2017.

% reference 36
\bibitem{WCNC2021} P. Ni, Z. Wang, H. Li, M. Li, and Q. Liu, ``Joint user scheduling and hybrid beamforming design for cooperative mmWave networks," in \textit{Proc. IEEE Wireless Commun. Networking Conf. (WCNC)}, Nanjing, China, Apr. 2021.


% channel eatimation
\bibitem{A. Alkhateeb}      A. Alkhateeb, O. E. Ayach, G. Leus, and R. W. Heath, ``Channel estimation and hybrid precoding for millimeter wave cellular systems," \textit{IEEE J. Sel. Topics Signal Process.}, vol. 8, no. 5, pp. 831-846, Oct. 2014.

% split the original problem
\bibitem{D. Zhao}     D. Zhao, H. Lu, Y. Wang, H. Sun, and Y. Gui, ``Joint power allocation and user association optimization for IRS-assisted mmWave systems,'' \textit{IEEE Trans. Wireless Commun.}, vol. 21, no. 1, pp. 577-590, Jan. 2022.

% matching theory
\bibitem{Y. Gu}      Y. Gu, \textit{et al.}, ``Matching theory for future wireless networks: Fundamentals and applications,"  \textit{IEEE Commun. Mag.}, vol. 53, no. 5, pp. 52-59, May 2015.

\bibitem{Z. Song}   Z. Song, Q. Ni, and X. Sun, ``Distributed power allocation for nonorthogonal multiple access heterogeneous networks," \textit{IEEE Commun.
    Lett.}, vol. 22, no. 3, pp. 622-625, Mar. 2018.

\bibitem{Y. Zhang2021}   Y. Zhang, L. Dai, and E. W. M. Wong, ``Optimal BS deployment and user association for 5G millimeter wave communication networks," \textit{IEEE Trans. Wireless Commun.}, vol. 20, no. 5, pp. 2776-2791, May 2021.

\bibitem{T. H. Cormen sort}   T. H. Cormen, C. E. Leiserson, R. L. Rivest, and C. Stein, ``Introduction to Algorithms", 3rd ed. Cambridge, MA, USA: MIT Press, 2009.

%%%% Algorithms
% FP 1
\bibitem{K. Shen1}         K. Shen and W. Yu, ``Fractional programming for communication systems-part I: Power control and beamforming," \textit{IEEE Trans. Signal Process.}, vol. 66, no. 10, pp. 2616-2630, May 2018.

% FP 2
\bibitem{K. Shen2}         K. Shen and W. Yu, ``Fractional programming for communication systems-part II: Uplink scheduling via matching," \textit{IEEE Trans. Signal Process.}, vol. 66, no. 10, pp. 2631-2644, May 2018.

% SDR  + QCQP
\bibitem{Z. Luo}  Z. Luo, W. Ma, A. M. So, Y. Ye, and S. Zhang, ``Semidefinite relaxation of quadratic optimization problems," \textit{IEEE Signal Process. Mag.}, vol. 27, no. 3, pp. 20-34, May 2010.

% 共轭梯度算法
\bibitem{L. Zhang}         L. Zhang, W. Zhou, and D. Li, ``A descent modified Polak-Ribiere-Polyak conjugate gradient method and its global convergence," \textit{IMA J. Numerical Anal.}, vol. 26, no. 4, pp. 629-640, Oct. 2006.

% 共轭梯度算法收敛性
\bibitem{P.-A. Absil}      P.-A. Absil, R. Mahony, and R. Sepulchre, ``Optimization algorithms on matrix manifolds,'' \textit{Princeton}, NJ, USA: Princeton Univ. Press, 2009.

% 动态天线
\bibitem{T. Kanungo}       T. Kanungo, D. M. Mount, N. S. Netanyahu, C. D. Piatko, R. Silverman, and A. Y. Wu, ``An efficient K-means clustering algorithm: Analysis and implementation," \textit{IEEE Trans. Pattern Anal. Mach. Intell.}, vol. 24, no. 7, pp. 881-892, Jul. 2002.

% 能量效率
\bibitem{R. Mendez-Rial}   R. M.-Rial, C. Rusu, N. G.-Prelcic, A. Alkhateeb, and R. W. Heath, ``Hybrid MIMO architectures for millimeter wave communications: Phase shifters or switches?" \textit{IEEE Access}, vol. 4, pp. 247-267, Jan. 2016.

\end{thebibliography}
\end{document}